\documentclass[entropy,article,submit,moreauthors,pdftex]{Definitions/mdpi} 
\usepackage{tikz}
\usetikzlibrary{positioning}
\usepackage{amsmath,amsfonts,amssymb,psfrag,amsthm}
\usepackage{graphicx}
\usepackage{epstopdf}
\usepackage{rotating}
\usepackage{dsfont}
\usepackage{color}
\usepackage{tikz}
\usetikzlibrary{plotmarks}
\usepackage{pgfplots}
\usetikzlibrary{calc}%,fillbetween
\usetikzlibrary{shapes,arrows}
\usetikzlibrary{decorations.markings}
\usetikzlibrary{positioning}
\pgfplotsset{compat=1.10}
\usetikzlibrary{calc}%,fillbetween
\usetikzlibrary{shapes,arrows}
\usetikzlibrary{decorations.markings}

\usepackage{eurosym}
\usepackage{algorithm}
\usepackage{algpseudocode}
\usepackage{caption}

\makeatletter
\newenvironment{breakablealgorithm}
{
    \begin{center}
    \refstepcounter{algorithm}
    \hrule height .8pt depth 0pt \kern 2pt
    \renewcommand{\caption}[2][\relax]{%
        {\raggedright
        \textbf{\ALG@name~\thealgorithm} ##2\par}%
        \ifx\relax##1\relax
        \addcontentsline{loa}{algorithm}
        {\protect\numberline{\thealgorithm}##2}%
        \else
        \addcontentsline{loa}{algorithm}
        {\protect\numberline{\thealgorithm}##1}%
        \fi
        \kern 2pt
        \hrule
        \kern 2pt
    }
}
{
    \kern 2pt
    \hrule
    \end{center}
}
\makeatother

\firstpage{1} 
\pubvolume{1}
\issuenum{1}
\articlenumber{0}
\pubyear{2026}
\copyrightyear{2026}
\datereceived{} 
\dateaccepted{} 
\datepublished{} 
\hreflink{https://doi.org/} % If needed use \linebreak

\Title{Bayesian-Guided Cooperative RL Beamforming for Wireless Adversarial User Detection}

\TitleCitation{Bayesian-Guided Cooperative RL Beamforming for Wireless Adversarial User Detection}

\Author{Parmida Geranmayeh$^{1,*}$\orcidA{} and Onur G\"unl\"u$^{1,2}$\orcidB{}}

\AuthorNames{Parmida Geranmayeh and Onur G\"unl\"u}

\AuthorCitation{Geranmayeh, P.; G\"unl\"u, O.}

\address{%
$^{1}$ \quad Lehrstuhl für Nachrichtentechnik, Technische Universität Dortmund, 44227 Dortmund, Germany; \{parmida.geranmayeh, onur.guenlue\}@tu-dortmund.de\\
$^{2}$ \quad Information Theory and Security Laboratory (ITSL), Link{\"o}ping University, 581 83 Linköping, Sweden\quad\\
}

\corres{Correspondence: parmida.geranmayeh@tu-dortmund.de.}

\conference{\citep{Geranmayeh2026ISAC}}

\abstract{ 
In next-generation wireless networks, communication systems are expected to go beyond simple data transmission and simultaneously provide high data rates, efficiency, and security. This requirement has motivated the extensive adoption of machine learning methods to develop intelligent and real-time network management frameworks, enabling the system to continuously monitor and react to channel variations and user behavior while maintaining efficient information delivery. In this context, the integration of machine learning with beamforming enables adaptive and data-driven beam direction selection, improving both the efficiency and security of wireless links. In this work, a 3GPP-based system model is first implemented under a no-attacker scenario, and an exhaustive search is employed as a reference to identify the best beamforming configurations. The proposed framework is then evaluated in the presence of an attacker and under different network scalability conditions. We demonstrate that the reinforcement learning-based approaches, namely Q-learning and SARSA (State-Action-Reward-State-Action), consistently outperform random selection in terms of total channel capacity, attacker detection accuracy, and performance stability. Among the evaluated reinforcement learning methods, Q-learning achieves the best overall trade-off between detection accuracy and computational efficiency. Our results indicate that the proposed framework provides a stable, scalable, and effective solution for joint beamforming and security-aware decision-making in dynamic and adversarial wireless environments.
}


\usepackage[utf8]{inputenc} 
\usepackage[T1]{fontenc}

\usepackage{dsfont}
\usepackage{units}

\usepackage{url}              % provides \url{...}

\usepackage{amsmath}  % Use the [cmex10] option to ensure complicance
\usepackage{graphicx}         % provides \includegraphics{...} to
\usepackage{enumitem}
\usepackage{booktabs}         % fixes poor spacing in 

\usepackage{amsmath,amsfonts,amssymb,psfrag}
\usepackage{epstopdf}
\usepackage{rotating}
\usepackage{dsfont}
\usepackage{color}
\usepackage{tikz}
\usetikzlibrary{plotmarks}
\usepackage{pgfplots}
\usetikzlibrary{calc}%,fillbetween
\usetikzlibrary{shapes,arrows}
\usetikzlibrary{decorations.markings}
\usetikzlibrary{positioning}
\pgfplotsset{compat=1.10}
\usetikzlibrary{calc}%,fillbetween
\usetikzlibrary{shapes,arrows}
\usetikzlibrary{decorations.markings}

\usepackage{balance}

\usepackage[utf8]{inputenc}
\usepackage{algorithm}
\usepackage{algpseudocode}
\usepackage{float}

\def\BibTeX{{\rm B\kern-.05em{\sc i\kern-.025em b}\kern-.08em T\kern-.1667em\lower.7ex\hbox{E}\kern-.125emX}}

\DeclareFontFamily{U}{mathx}{\hyphenchar\font45}
\DeclareFontShape{U}{mathx}{m}{n}{<-> mathx10}{}
\DeclareSymbolFont{mathx}{U}{mathx}{m}{n}

\usepackage{scalerel}
\usepackage{stackengine}
\stackMath
\def\hatgap{2pt}
\def\subdown{-2pt}
\newcommand\reallywidehat[2][]{%
	\renewcommand\stackalignment{l}%
	\stackon[\hatgap]{#2}{%
		\stretchto{%
			\scalerel*[\widthof{$#2$}]{\kern-.6pt\bigwedge\kern-.6pt}%
			{\rule[-5\textheight]{0.1ex}{\textheight}}%WIDTH-LIMITED BIG WEDGE
		}{0.5ex}% THIS SQUEEZES THE WEDGE TO 0.5ex HEIGHT
		_{\smash{\belowbaseline[\subdown]{\scriptstyle#1}}}%
}}

\usepackage{enumitem}
\setlist[itemize]{nosep, left=1.5em} % Adjust 'left' as needed

\begin{document}

\section{Introduction}\label{sec:introduction}

In recent years, the integration of sensing and communication has emerged as one of the most promising paradigms for future wireless networks, attracting attention from both academia and industry \cite{Ding2022DigitalTwin,Geok2018RayTracing}. By combining sensing and communication functionalities within a unified infrastructure, Integrated Sensing and Communication (ISAC) is widely regarded as a key enabling technology for sixth-generation (6G) networks \cite{OnurRoleofISAC}. Extensive research has focused on waveform design, resource allocation, beamforming, and system performance optimization within this paradigm \cite{Sheen2020DigitalTwinRIS,Jalali2023SWIPT}.
Among these technologies, beamforming plays a fundamental role in improving communication performance by directing signal energy toward intended users while suppressing adversarial interference \cite{Bakhtiar2023Beamforming,Chary2024AIBeamforming, Geranmayeh2022BeamSelection}. Consequently, a wide range of beamforming techniques has been proposed to enhance the efficiency of next-generation wireless systems. For example, Ihsan et al. \cite{Ihsan2022IRS} integrated Intelligent Reflective Surfaces (IRS) with non-orthogonal multiple access (NOMA)-based beamforming to jointly optimize beamforming and power allocation for improved energy efficiency. Jiang et al. \cite{Jiang2022ICC} proposed a complementary-beam initialization strategy to achieve more uniform spatial coverage, while Qi et al. \cite{Qi2021UAVBeamforming} developed an Unmanned Aerial Vehicle (UAV)-assisted IRS architecture for Internet-of-Things (IoT) applications, demonstrating the effectiveness of beamforming in extending communication range and enhancing backscatter communication performance.

As wireless networks continue to increase in scale and complexity, beamforming has increasingly been incorporated into digital twin frameworks to support more accurate network modeling and optimization \cite{Geranmayeh2022BeamSelection}. The effectiveness of such frameworks, however, largely depends on the accuracy of the underlying wireless channel model. Consequently, ray tracing has become a key technique for channel modeling and wave propagation analysis \cite{Geok2018RayTracing,Ji2021RayTracingIndoor,Hsiao2017RayTracing}. By accurately modeling propagation paths, reflections, refractions, and diffractions, ray tracing provides realistic channel information that improves beamforming optimization and network performance, making it a valuable component of digital twin-enabled wireless systems \cite{Geranmayeh2022BeamSelection}.

The importance of realistic propagation modeling has been demonstrated in numerous studies. Ji et al. \cite{Ji2021RayTracingIndoor} proposed an advanced propagation prediction method for ultra-high frequency (UHF) environments using irregular spatial partitioning to improve modeling accuracy. Hsiao et al. \cite{Hsiao2017RayTracing} investigated millimeter-wave (mmWave) propagation under both line-of-sight (LOS) and non-line-of-sight (NLOS) conditions. A comprehensive survey presented in \cite{Yun2015RayTracingTheory} reviewed ray-tracing algorithms and their applications in wireless channel modeling, while Tiberi et al. \cite{Tiberi2009UWB} analyzed ultra-wideband channel characteristics through ray-tracing simulations. These studies confirm that accurate propagation modeling is indispensable for the design, evaluation, and optimization of future wireless communication systems. 

Besides channel propagation, antenna characteristics also play an essential role in beamforming performance \cite{Jalali2023SWIPT,Bakhtiar2023Beamforming}. Radiation patterns, antenna gain, polarization, and frequency response directly influence how electromagnetic energy propagates through the wireless environment and is received by users. Therefore, realistic antenna modeling complements ray-tracing-based channel characterization and contributes to more accurate beamforming optimization.

To efficiently optimize beamforming in dynamic wireless environments, reinforcement learning (RL) has emerged as a powerful alternative to exhaustive search methods \cite{Geranmayeh2025DQN}. Unlike exhaustive search, whose computational complexity increases rapidly with network size, RL continuously interacts with the environment to learn adaptive beam selection policies while maintaining computational efficiency. This online learning capability makes RL particularly suitable for highly dynamic communication scenarios in which channel conditions change over time. Furthermore, deep RL (DRL) has been successfully applied to resource allocation and network optimization problems in modern communication systems experiencing rapidly increasing traffic demands \cite{Vivekanand2024DRL}.

While communication efficiency remains a primary objective, ensuring network security has become equally important in next-generation wireless systems \cite{OnurSecureISACJournal,OnurSecurityTutorial, Zhou2026FeedbackLunch,Onur2026secureISACTutorial}. Game-theoretic approaches have therefore been widely employed to model strategic interactions between legitimate users and malicious entities under uncertainty. Lei et al. \cite{Lei2018Game} introduced a Markov game framework with incomplete information in which attack and defense strategies evolve dynamically. Jiang et al. \cite{Jiang2021Game} proposed a multi-stage signaling game for Moving Target Defense (MTD), whereas Teófilo et al. \cite{Teofilo2012Poker} illustrated the importance of adaptive decision-making in uncertain communication environments. These studies demonstrate that game theory provides an effective framework for modeling attacker–defender interactions in complex wireless networks.

Moreover, security challenges become even more critical in covert communication scenarios, where interference, noise, and jamming simultaneously affect communication reliability and detectability \cite{Yu2023Covert}. In addition to intentional jamming, fading, multipath propagation, and temporal channel variations introduce random fluctuations in signal amplitude and phase, making the distinction between legitimate and malicious transmissions considerably more difficult \cite{Esmaili2026Covert}. Consequently, traditional analytical approaches often struggle to cope with such highly dynamic environments.

To address these challenges, data-driven techniques have attracted increasing attention. Machine learning methods provide powerful tools for modeling the nonlinear and time-varying behavior of wireless channels, frequently outperforming conventional analytical techniques in complex environments \cite{Elsadig2022Covert}. Recent research has therefore focused on integrating machine learning, RL, and game-theoretic concepts. Ren et al. \cite{Ren2023UAV} proposed a hybrid framework combining incomplete-information games with DRL to derive Bayesian Nash equilibrium strategies for UAV networks. Yao et al. \cite{Yao2024BayesianGame} introduced a Bayesian Q-learning approach for cyber defense in cyber-physical systems. Similarly, \cite{Mamaghani2026ISAC} presented a secure ISAC framework that combines Stackelberg game theory and DRL to jointly optimize security, sensing performance, and energy efficiency. With the increasing density of spectrum access in 5G New Radio (NR) systems, Bayesian inference has also emerged as an effective tool for detecting jamming and subtle interference patterns \cite{Jere2023BayesianJamming}. Furthermore, trust-based detection mechanisms employing active learning and Bayesian classifiers have demonstrated promising capabilities for identifying malicious behaviors in wireless communication networks \cite{Huang2022VANET}.

Despite recent advances, existing studies have largely investigated beamforming optimization, channel modeling, security mechanisms, and Bayesian-based attacker detection independently. To address this gap, this work proposes an integrated framework that combines ray-tracing-based channel modeling, game-theoretic receiver (RX) association, RL-based beamforming, and Bayesian belief updating for secure beam management. By jointly integrating these techniques, the proposed framework improves total channel capacity, significantly enhances attacker detection accuracy, and provides a robust and scalable solution for intelligent and security-aware beamforming in dynamic wireless environments.

\subsection{Main Contributions}
The main contributions of this work are summarized as follows:

\begin{itemize}

\item A comprehensive intelligent framework is proposed for secure beamforming in 5G mmWave networks by integrating 3GPP TR~38.901-based channel modeling, deterministic ray tracing, Bayesian-guided RL, and game-theoretic user association within a unified decision-making architecture. The Bayesian belief idea is related to~\cite{Geranmayeh2026ISAC}. However, \cite{Geranmayeh2026ISAC} infers a single attacker from variations in the aggregate network capacity, whereas the present work forms link-specific evidence, supports multiple attackers, and incorporates the resulting beliefs into cooperative transmitter beam selection, enabling more accurate attacker localization and improved detection performance.

\item A scalable Bayesian-guided cooperative RL framework is developed and evaluated under multiple network configurations with different RX densities (with RX antenna numbers $\in\{4,6,8\}$) and varying numbers of malicious users, whereas \cite{Geranmayeh2026ISAC} considers only a single 2-TX/4-RX, single-attacker configuration.

\item A cooperative beamforming strategy is introduced, in which two transmitters (TXs) jointly determine their beamforming decisions instead of operating independently. This cooperative decision-making is shown to improve the coordination between TXs and enhance both communication performance and attacker mitigation.

\item An exhaustive search-based beamforming benchmark is introduced to determine the best beam steering angles for all TXs and RXs under the baseline scenario. The proposed framework identifies and reports the best cooperative beam configurations for all network nodes, providing a reliable reference for evaluating the learning-based methods.

\item The RL framework is substantially extended through a comprehensive comparison of Q-learning, State-Action-Reward-State-Action (SARSA), and random beam selection. Under the evaluated network configurations, the results show that the RL-based methods achieve higher throughput and attacker-detection accuracy than random beam selection. Among the evaluated RL methods, Q-learning provides the most favorable overall performance across the considered communication, detection, and runtime metrics.

\item Extensive Monte Carlo simulations are conducted for
$N_{\mathrm{RX}}\in\{4,6,8\}$ to evaluate the empirical performance of the proposed framework under different network sizes and adversarial conditions.

\end{itemize}

\subsection{Paper Organization}
In Section \ref{sec:SystemModelandNetworkArchitecture}, the system model and simulation setup are presented, including the 3GPP Urban Micro-cellular (UMi) scenario, the considered geometric configuration (i.e., Dortmund urban environment in Germany with TX and RX placement), the beamforming formulation, and exhaustive search for beamforming for all antennas. Section \ref{sec:PROPOSEDMETHODOLOGY} presents the implementation framework, including the RL formulations and the Bayesian-based utility function employed for attacker detection. Section \ref{sec:SimulationandDiscussions} provides a comprehensive performance evaluation and comparison of the proposed approaches. Finally, Section \ref{sec:conclusions} summarizes the main conclusions of the study and outlines potential directions for future work.

\section{System Model and Network Architecture}\label{sec:SystemModelandNetworkArchitecture}
Consider a 5G NR wireless communication system operating in the mmWave frequency band. The network follows a 3GPP-inspired UMi-type deployment. The numbers of TXs and RXs are fixed as $N_{\mathrm{TX}}=2$ and $N_{\mathrm{RX}}\in\{4,6,8\}$, while their locations are sampled uniformly within the considered deployment area, equivalently corresponding to a PPP conditioned on the specified numbers of nodes. To evaluate the proposed framework under adversarial conditions, one or two RX-associated user nodes are randomly designated as malicious users (attackers), i.e., $N_{\mathrm{Attacker}}\in\{1,2\}$. 

The one- and two-attacker settings provide a controlled transition from a single-attacker case to a multi-attacker case while preserving a meaningful number of legitimate RXs in the evaluated networks. This choice is an evaluation constraint rather than an inherent restriction of the Bayesian belief formulation.

Uniform rectangular array (URA) antennas are employed to enable directional beamforming and MIMO operation. A $4\times 8$ URA, i.e., 32 antenna elements, is considered on the TX side, and a $2\times 2$ URA, i.e., 4 antenna elements, is considered on the RX side. The carrier frequency is set to $f_c=28\,\mathrm{GHz}$, which lies in the 5G NR FR2 range, and the system operates over a bandwidth of $B=400\,\mathrm{MHz}$.

The network geometry is designed such that the height of the Base Stations (BS) ranges between $ 8$ and $ 23$ meters randomly, and the height of the User Equipments (UE) is randomly determined between $1.2$ and $2$ meters. Based on standard constraints, the minimum allowable distance from a user to the nearest base station is set to $10$ meters, and the Inter-Site Distance (ISD) is configured to be $200$ meters. The communication links between nodes may experience either line-of-sight (LOS) or non-line-of-sight (NLOS) propagation conditions. The probability that a link between a TX and an RX is in the LOS state at a three-dimensional distance $d$ is determined according to the 3GPP TR~38.901 UMi model as

\begin{equation}
P_{\mathrm{LOS}}(d)=\min\left(\frac{18}{d},1\right)\left(1-\exp^{-\frac{d}{36}}\right)+\exp^{-\frac{d}{36}}.
\end{equation}

If the link is in the LOS state, the signal propagates over a direct and unobstructed path. Otherwise, the link is considered to be in the NLOS state, where the received signal is formed through reflections, diffractions, and scattering from surrounding objects. Accordingly, the large-scale path loss is computed using the corresponding 3GPP TR~38.901 UMi path-loss model \cite{ITU_R_M2135_1}:

\begin{align}
PL_{\mathrm{LOS}}(d)&=32.4+21\log_{10}(d)+20\log_{10}(f_c),\\
PL_{\mathrm{NLOS}}(d)&=32.4+31.9\log_{10}(d)+20\log_{10}(f_c).
\end{align}

where $f_c$ denotes the carrier frequency and is set to $28\,\mathrm{GHz}$. The larger distance-dependent coefficient in the NLOS model reflects the additional attenuation caused by buildings and other urban obstructions, resulting in more severe propagation loss under blocked transmission conditions.

\subsection{Spatial Topology and Node Deployment}
To accurately model the spatial distribution of the nodes, a PPP with a specified density is utilized. After generating the random positions of TXs and RXs in a two-dimensional Cartesian space bounded by the interval $[-150, 150]$ meters, the network coordinates are mapped onto geographical coordinates (latitude and longitude) based on the reference point $(51.5139^{\circ}\text{N}, 7.4653^{\circ}\text{E})$ corresponding to Dortmund Downtown. This mapping aligns the setup with real-world urban telecommunication maps extracted from the OpenStreetMap (OSM) database. The three-dimensional position of the nodes in meters within the Cartesian simulation structure is formulated as follows and applied to the network geometry matrix:

\begin{align}
    &\mathbf{p}_{\text{TX}, i} = [x_{\text{TX}, i}, y_{\text{TX}, i}, z_{\text{TX}, i}]^T, \quad i \in \{1, 2\},\\
    &\mathbf{p}_{\text{RX}, j} = [x_{\text{RX}, j}, y_{\text{RX}, j}, z_{\text{RX}, j}]^T, \quad j \in \{1, \dots, N_{\text{RX}}\}
\end{align}

\begin{figure}[t]
	\centering
\includegraphics[width=0.95\columnwidth, keepaspectratio]{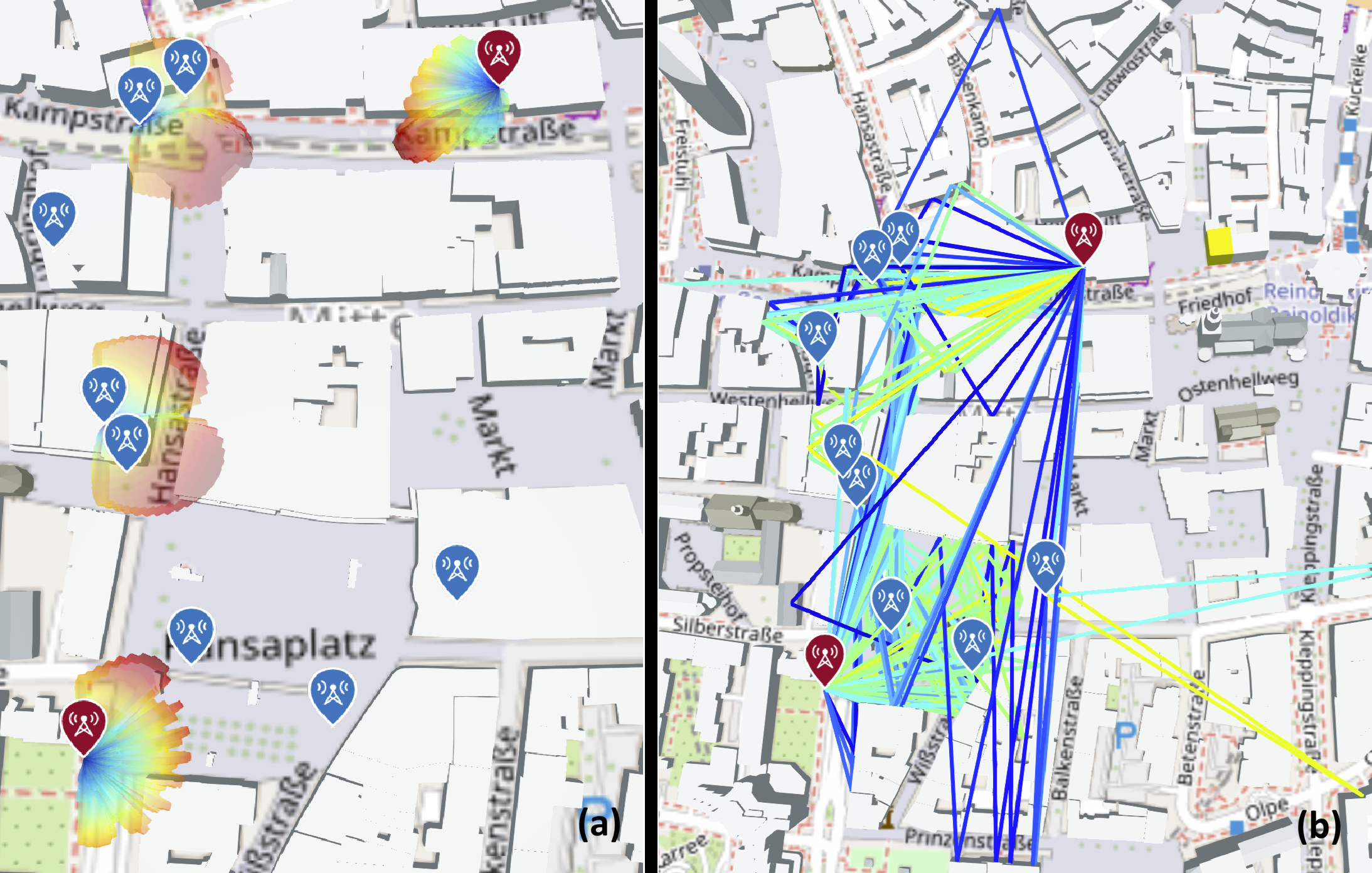}
	\caption{Ray-tracing environment and network topology for 2 TXs and 8 RXs.}
	\label{fig:city view}
\end{figure}

Table~\ref{tab:node_coordinates} presents the node coordinates for a representative network realization, and the corresponding geometric configuration is illustrated in  Figure~\ref{fig:city view}. To improve visual clarity, Figure~\ref{fig:city view}(a) illustrates the beamforming patterns of the two TXs toward two representative RXs. Figure~\ref{fig:city view}(b) presents the locations of all TXs and RXs within the considered deployment area with all ray-tracing propagation paths.

\begin{table}[ht]
\centering
\caption{Three-dimensional coordinates of the network nodes.}
\label{tab:node_coordinates}
\begin{tabular}{cccc}
\toprule
\textbf{Node} & \textbf{Height (m)} & \textbf{Latitude ($^\circ$)} & \textbf{Longitude ($^\circ$)} \\
\midrule
TX1 & 9.6815  & 51.512563 & 7.463546 \\
TX2 & 12.6455 & 51.515124 & 7.465976 \\
RX1 & 1.2863  & 51.513811 & 7.463668 \\
RX2 & 1.4439  & 51.514514 & 7.463356 \\
RX3 & 1.5096  & 51.513119 & 7.465642 \\
RX4 & 1.7802  & 51.513637 & 7.463765 \\
RX5 & 1.9954  & 51.515053 & 7.463836 \\
RX6 & 1.7769  & 51.512653 & 7.464935 \\
RX7 & 1.2875  & 51.515169 & 7.464115 \\
RX8 & 1.5010  & 51.512890 & 7.464137 \\
\bottomrule
\end{tabular}
\end{table}

\subsection{Ray-Tracing and Deterministic Channel Model}
To provide a realistic representation of mmWave propagation, a deterministic channel model based on the Ray-Tracing Shooting and Bouncing Rays (SBR) algorithm is employed. Unlike simplified statistical channel models, the adopted approach explicitly accounts for the physical propagation environment, including building geometry, urban obstacles, and realistic concrete material properties. For each TX--RX link, the wireless channel is represented by a multipath Channel Impulse Response (CIR). The ray-tracing simulation considers at most one reflection and one diffraction for each propagation path. The channel parameters extracted from each propagation ray include the path loss $PL_k$, propagation phase $\phi_k$, Angle of Departure (AoD), and Angle of Arrival (AoA) in both the azimuth and elevation domains.

The network geometry, deployment assumptions, and LOS/NLOS conditions follow the standard 3GPP TR~38.901 UMi specification. Accordingly, the corresponding 3GPP LOS probability and path-loss expressions are presented as a reference propagation model. However, in the proposed simulations, the actual path loss is obtained directly from the SBR ray-tracing engine rather than being computed using the analytical 3GPP path-loss equations. Consequently, the 3GPP path-loss model serves only as a reference for comparison, whereas all performance evaluations are conducted using the ray-tracing-derived channel parameters.

\subsection{Spatially Correlated Shadow Fading Model}
Shadow fading (or slow fading) represents the slow variations in received power caused by large environmental obstacles. In this system, the shadowing effect is modeled as a spatially correlated Gaussian random field, which is added to the path loss on a dB scale. The standard deviation of this log-normal distribution is set to $\sigma_{\text{shadow}} = 7 \text{ dB}$, and its environmental correlation distance is $d_{\text{corr}} = 50 \text{ m}$. The structure of the correlated field is generated by applying a two-dimensional convolution over an independent and identically distributed (i.i.d.) Gaussian matrix using an exponential correlation kernel based on the relative physical distance grid step $dx$ such that \cite{Gudmundson1991Correlation}

\begin{equation}
    K(\Delta x) = \exp\left( -\frac{|\Delta x| \cdot dx}{d_{\text{corr}}} \right).
\end{equation}

The spatial filtering kernel is normalized such that $\sum K = 1$. Moreover, the shadowing attenuation for the link between TX $i$ and RX $j$ is extracted via a two-dimensional interpolation of the field values at the respective positions and is applied as the sum of shadowing effects at both ends of the link (i.e., $s_{ij} = s_{\text{TX}, i} + s_{\text{RX}, j}$).

\subsection{Directional Antenna and Discrete Beamforming Model}
To compensate for the severe path loss at the $28\text{ GHz}$ frequency band, all TXs and RXs are equipped with microstrip patch directional antenna arrays. The antenna radiation characteristics are incorporated into the simulation environment through a pre-computed beam pattern dataset (\texttt{AntennaPattern.mat}), which was generated offline using the adopted antenna array model and stores the angle-dependent radiation responses of the antenna array.

The beamforming and positioning framework is based on a discrete angular codebook. Specifically, the beam steering directions are uniformly quantized over the interval $[-60^\circ, 60^\circ]$ with a step size of $5^\circ$, resulting in $N_{\text{angle}} = 25$ discrete beam directions for each antenna, defined as:

\begin{equation}
A = \{-60^\circ,\,-55^\circ,\,\ldots,\,0^\circ,\,\ldots,\,55^\circ,\,60^\circ\}.
\end{equation}

The interval $[-60^\circ,60^\circ]$ represents the useful scanning sector of the adopted directional antenna arrays. Steering toward directions close to end-fire generally reduces the effective aperture and array gain and increases beam distortion and sidelobe sensitivity. The $5^\circ$ spacing provides a compromise between angular resolution and learning complexity, resulting in 25 candidate directions per antenna. Extending the angular interval or adopting a finer resolution would enlarge the joint action space and increase the convergence time of the tabular RL methods.

\begin{figure}[t]
    \centering
    \includegraphics[width=0.70\columnwidth,keepaspectratio]{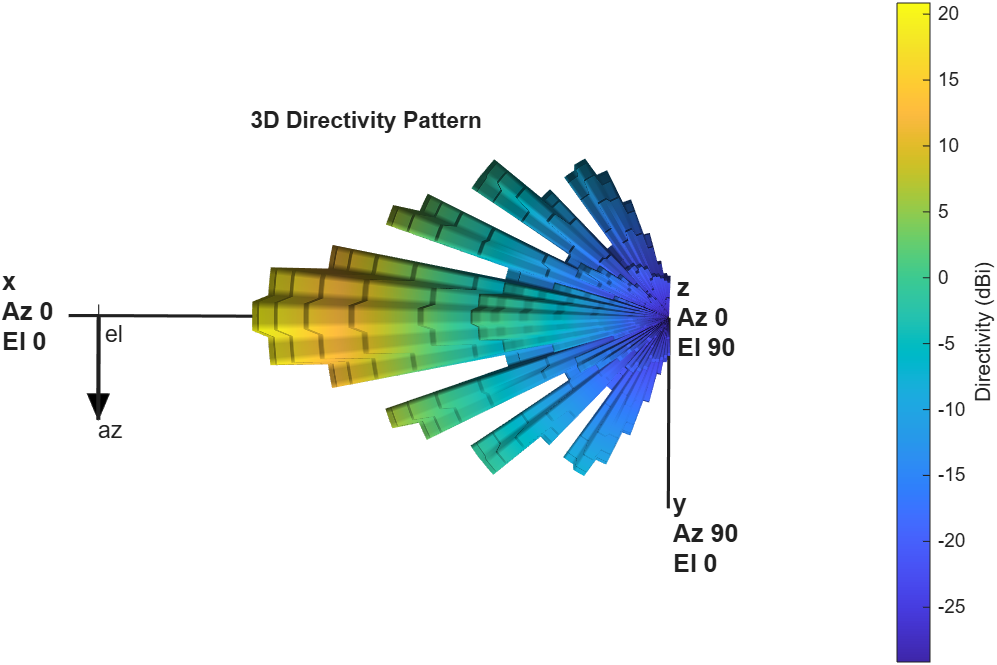}
    \caption{\protect Digitized radiation pattern of the adopted
    directional antenna array over the considered beam-steering sector.}
    \label{fig:digitized_antenna_radiation_pattern}
\end{figure}

Furthermore, the inter-element spacing for the URA antenna structures is set to half of the carrier wavelength. Beam steering is achieved using array steering vectors, which define phase shifts across antenna elements such that constructive interference is formed in the desired direction while suppressing radiation in other directions. Accordingly, the beamforming weight vector is defined as:

\begin{equation}
\mathbf{w} = \mathbf{a}(f,\theta),
\end{equation}

where $\mathbf{a}(f,\theta)$ denotes the steering vector corresponding to frequency $f$ and angle $\theta$. After determining the beamforming weight vector, the array radiation pattern for each steering direction is obtained by applying these weights to the antenna elements, generating the corresponding directional beam. The effective communication link is then characterized by incorporating beamforming at both TX and RX sides. Specifically, the effective channel gain is computed as:

\begin{equation}
G = \left|\mathbf{w}_{\mathrm{r}}^{H}\mathbf{H}\mathbf{w}_{\mathrm{t}}\right|^{2},
\label{eq:effective_channel_gain}
\end{equation}

where $\mathbf{H}$ represents the channel matrix, $\mathbf{w}_t$ is the transmit beamforming vector, and $\mathbf{w}_r$ is the receive beamforming vector. This formulation captures the degree of alignment between the TX–RX beam directions and the underlying multipath channel structure.  Figure~\ref{fig:digitized_antenna_radiation_pattern} illustrates the resulting TX beam pattern.

\subsection{Game-Theoretic Node Assignment Mechanism}
The association and assignment of RXs to TXs during the network initialization phase are modeled as a non-cooperative game with stochastic best-response dynamics. In this framework, each RX sequentially updates its association decision based on instantaneous effective link gains, interference conditions, and load-dependent congestion effects. The proposed mechanism does not assume centralized coordination and operates in a fully distributed manner.

The utility function for RX $j$ when associated with TX $i$ is defined as:

\begin{equation}
U_j(i) = \log_{2}(1 + \mathrm{SINR}_j) - \beta \log(N_i + 1),
\end{equation}

where $N_i$ represents the number of RXs currently associated with TX $i$, introducing a congestion-aware penalty term to mitigate load imbalance. The parameter $\beta = 0.8$ controls the impact of this load balancing effect. The effective link gain is computed using a ray-tracing-based channel model combined with spatial shadow fading and antenna pattern responses, and is used to evaluate the instantaneous SINR during each association update.

User association decisions are updated iteratively using a Softmax-based probabilistic best-response rule with temperature parameter $\tau = 0.7$, given by:

\begin{equation}
P_j(i) =
\frac{\exp(U_j(i)/\tau)}
{\sum_{m=1}^{N_{\mathrm{TX}}}\exp(U_j(m)/\tau)}.
\end{equation}

The association process is executed over a finite number of iterations (up to 15 iterations), leading to a stable and stationary association configuration under the adopted stochastic response dynamics. All hyperparameters employed in the proposed framework, including the game-theoretic, Bayesian, and RL parameters, were determined experimentally based on extensive preliminary simulations.

\subsection{Interference, Attacker, and SINR Model}

The thermal-noise power spectral density is given by
$N_0=-174~\mathrm{dBm/Hz}$. Considering the system bandwidth
$B=400~\mathrm{MHz}$ and the RX noise figure
$NF_{\mathrm{UE}}=13~\mathrm{dB}$, the total noise power is computed
as \cite{ETSI_TR_138901_v1610}

\begin{align}
N_{\mathrm{dBm}}
&=
N_0 + 10\log_{10}(B) + NF_{\mathrm{UE}},
\\
N_{\mathrm{W}}
&=
10^{\frac{N_{\mathrm{dBm}}-30}{10}},
\end{align}

where $B$ is expressed in Hz when calculating the noise power.

The signal-to-interference-plus-noise ratio (SINR) for RX $j$, associated
with TX $\mathcal{C}(j)$, is defined as

\begin{equation}
\mathrm{SINR}_j
=
\frac{P_{\mathrm{signal}}}
{P_{\mathrm{interference}}+P_{\mathrm{noise}}}.
\end{equation}

Here, $P_{\mathrm{signal}}$ denotes the received power of the desired link, $P_{\mathrm{interference}}$ is the aggregate received interference power from unintended transmissions and malicious interference, and $P_{\mathrm{noise}}$ denotes the receiver thermal-noise power. A malicious RX deliberately selects an unfavorable receive beam that reduces the desired-link gain and/or increases the received interference, thereby decreasing its SINR and achievable capacity.
Consequently, this adversarial beam selection also reduces the
aggregate network capacity. Accordingly, the capacity of RX $j$ and the aggregate network capacity are given by

\begin{align}
C_j
&=
B\log_2\left(1+\mathrm{SINR}_j\right),
\\
C_{\mathrm{Total}}
&=
\sum_{j=1}^{N_{\mathrm{RX}}} C_j.
\label{eq:channelcapacity}
\end{align}

\section{PROPOSED METHODOLOGY}\label{sec:PROPOSEDMETHODOLOGY}

We next present the proposed intelligent beamforming framework, adaptive decision-making procedure, and threat management method within the considered mmWave network. The objective of this phase extends beyond simple beam angle selection; it aims to jointly optimize network capacity, mitigate interference, detect malicious users, counter jamming attacks, and continuously adapt to dynamic radio environments. The proposed framework combines RL and Bayesian inference within a unified beamforming architecture. This enables the learning agent to continuously adapt its decisions according to both the wireless channel conditions and the estimated likelihood of users being malicious.

\begin{figure*}[!t]
\centering

\begin{tikzpicture}[
    node distance=6mm,
    block/.style={
        rectangle,
        rounded corners=2pt,
        draw=black,
        thick,
        fill=blue!8,
        text width=7.5cm,
        minimum height=9mm,
        align=center,
        font=\small
    },
    flow/.style={
        ->,
        >=stealth,
        thick
    },
    feedback/.style={
        ->,
        >=stealth,
        thick,
        dashed
    }
]

\node[block] (A)
{Urban network and node deployment};

\node[block, below=of A] (B)
{Ray tracing and spatially correlated shadow fading};

\node[block, below=of B] (C)
{Game-theoretic TX--RX association};

\node[block, below=of C] (D)
{Attacker interference and link-level observations};

\node[block, below=of D] (E)
{Bayesian evidence and attacker-belief update};

\node[block, below=of E] (F)
{RL state construction and cooperative beam selection};

\node[block, below=of F] (G)
{Capacity evaluation and attacker detection};

\draw[flow] (A) -- (B);
\draw[flow] (B) -- (C);
\draw[flow] (C) -- (D);
\draw[flow] (D) -- (E);
\draw[flow] (E) -- (F);
\draw[flow] (F) -- (G);

% Feedback loop
\draw[feedback]
    (G.east)
    -- ++(1.2,0)
    |- (D.east);

\node[
    align=left,
    font=\scriptsize,
    right=15mm of E
]
{Performance\\feedback};

\end{tikzpicture}

\caption{Overall architecture of the proposed Bayesian-guided cooperative RL framework. Ray-tracing channel parameters and spatial shadow fading define the communication environment, game-theoretic association assigns RXs to TXs, and link-level observations update the RX-specific suspicion scores.
These scores are incorporated into the RL state used by the
two cooperative TXs for beam selection.}

\label{fig:proposed_architecture}
\end{figure*}
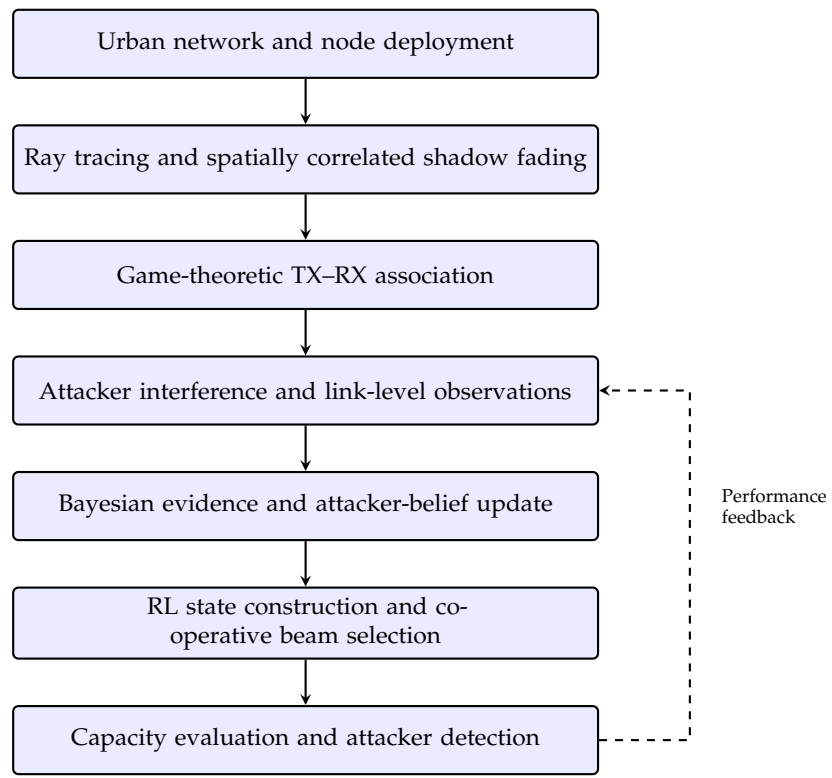

Figure ~\ref{fig:proposed_architecture} presents the overall architecture of the proposed framework. The propagation environment is first constructed using the urban geometry, ray-tracing channel parameters, and spatially correlated shadow fading. The RXs are then associated with the TXs through the game-theoretic allocation mechanism. During learning, link-level observations are used to update the RX-specific suspicion scores, which form part of the RL state. Based on this state, the two cooperative TXs select their beam directions using Q-learning or SARSA. The resulting communication performance is subsequently used to update both the Bayesian evidence and the learning policy.

At the beginning of each scenario, the positions of TXs and RXs are initialized, as described above, and spatially correlated shadow fading effects are extracted from a pre-generated shadow map. Subsequently, a single-run ray tracing procedure is performed for all TX-RX links, and the resulting channel parameters, including path loss, phase, AoA, and AoD, are stored. This precomputation significantly reduces computational complexity during the learning process, as it eliminates the need to repeatedly execute ray tracing within the RL loops. After constructing the communication environment, each RX is associated with a TX using a game-theoretic allocation mechanism. This mapping considers link quality, interference levels, and network load, thereby establishing the initial communication structure for subsequent decision-making.

Ray tracing and RL perform complementary rather than interchangeable functions in the proposed framework. Ray tracing constructs the site-specific propagation environment and provides the path loss, propagation phase, AoD, and AoA of each TX--RX link. Q-learning and SARSA subsequently use these stored channel parameters to select cooperative beam directions. Although this procedure introduces an offline channel-generation and training cost, the converged policy can select beam directions directly from the learned action values without performing a new exhaustive search.

The proposed problem involves several interdependent challenges. First, the joint beam-selection space grows exponentially with the number of TXs and RXs, making repeated exhaustive search impractical in larger network configurations. Second, the identities of the malicious RXs are not observable by the learning agent and must be inferred from link-level performance variations under interference and channel uncertainty. Third, increasing the number of RXs enlarges the learning space and makes the attribution of performance degradation to individual nodes more difficult.
To address these challenges, link-specific Bayesian evidence is used to estimate the likelihood of each RX being malicious, while cooperative RL enables the two TXs to adapt their beam directions without repeatedly evaluating the complete joint beam space.

In each Monte Carlo realization, a subset of RXs is randomly selected as malicious nodes (i.e., attackers). These nodes degrade network performance by introducing interference and disrupting normal communication. Since the identities of the attackers are unknown to the learning agent, the system iteratively assigns an RX-specific suspicion score to each node
based on its observed and counterfactual communication performance. These scores are collected in the vector

\begin{equation}
\mathbf{b}=[b_1,b_2,\ldots,b_{N_{\mathrm{RX}}}]^{T}, 
\qquad 0\leq b_i\leq 1,\quad i=1,\ldots,N_{\mathrm{RX}}.
\end{equation}

For each RX $i$, let $C_{i,\mathrm{exp}}^{(t)}$ denote the
counterfactual link capacity predicted under the current TX and nominal RX beam configuration in the absence of malicious interference, and let $C_{i,\mathrm{obs}}^{(t)}$ denote the capacity observed after applying the attacker-induced interference and beam perturbation. Let
$C_{\mathrm{exp}}^{(t)}=\sum_i C_{i,\mathrm{exp}}^{(t)}$ and
$C_{\mathrm{obs}}^{(t)}=\sum_i C_{i,\mathrm{obs}}^{(t)}$.

To obtain link-specific evidence, RX $i$ is counterfactually restored to its nominal beam while all other observed quantities are kept fixed.
Let $C_{\mathrm{rest},i}^{(t)}$ denote the resulting aggregate capacity. The local normalized capacity drop and counterfactual impact are defined as

\begin{align}
d_i^{(t)}
&=
\frac{
\left[C_{i,\mathrm{exp}}^{(t)}
-C_{i,\mathrm{obs}}^{(t)}\right]^+
}{
C_{i,\mathrm{exp}}^{(t)}+\delta
},\\
q_i^{(t)}
&=
\frac{
\left[C_{\mathrm{rest},i}^{(t)}
-C_{\mathrm{obs}}^{(t)}\right]^+
}{
C_{\mathrm{exp}}^{(t)}+\delta
}.
\end{align}

Here, $[x]^+=\max(0,x)$ and $\delta=10^{-12}$ is used to avoid division by zero. The instantaneous evidence score is

\begin{equation}
S_i^{(t)}
=
\min\left\{
1,\,
0.65q_i^{(t)}+0.35d_i^{(t)}
\right\}.
\end{equation}

The weights $0.65$ and $0.35$ were selected empirically to assign greater importance to the counterfactual network-level impact while retaining the RX-specific local capacity drop. They were fixed in all reported experiments and were not optimized separately for individual network scenarios. The evidence is temporally smoothed according to

\begin{equation}
E_i^{(t)}
=
\rho E_i^{(t-1)}
+
(1-\rho)S_i^{(t)},
\qquad
\rho=0.85,
\end{equation}

with $E_i^{(0)}=0$. The initial RX scores are set uniformly as

\begin{equation}
b_i^{(0)}=\frac{1}{N_{\mathrm{RX}}}.
\end{equation}

The smoothed evidence is mapped to the following bounded evidence multiplier:

\begin{equation}
\Lambda_i^{(t)}
=
\exp\!\left(
\operatorname{clip}
\left(
6(E_i^{(t)}-0.02),-4,4
\right)
\right),
\end{equation}

where
$\operatorname{clip}(x,l,u)=\min\{u,\max\{l,x\}\}$. During the initial burn-in period of 300 iterations, the scores remain at their initial values. Subsequently, the RX-specific suspicion score is updated in odds form as

\begin{equation}
b_i^{(t)}
=
\frac{
b_i^{(t-1)}\Lambda_i^{(t)}
}{
1-b_i^{(t-1)}
+
b_i^{(t-1)}\Lambda_i^{(t)}
}.
\end{equation}

The multiplier $\Lambda_i^{(t)}$ is a deterministic calibration of the empirical evidence and is not estimated from class-conditional probability densities. Therefore, $b_i^{(t)}$ is conservatively referred to as an RX-specific suspicion score rather than a rigorously calibrated Bayesian posterior probability. The scores are updated independently and are not normalized to sum to one, thereby allowing multiple RXs to attain large suspicion scores simultaneously.

At iteration $t$, the most suspicious RX is

\begin{equation}
i^{*(t)}=\arg\max_i b_i^{(t)}.
\end{equation}

Its score is quantized into five confidence levels according to

\begin{equation}
L^{(t)}
=
\min\left\{
5,\,
\left\lfloor5b_{i^{*(t)}}^{(t)}\right\rfloor+1
\right\}.
\end{equation}

Thus, $L=1,2,3,4,5$ correspond respectively to
$b_{i^*}\in[0,0.2)$, $[0.2,0.4)$, $[0.4,0.6)$, $[0.6,0.8)$, and
$[0.8,1]$. The tabular RL state is
$s^{(t)}=(i^{*(t)},L^{(t)})$. For final detection, the
$N_{\mathrm{Attacker}}$ RXs with the largest scores are returned as the estimated attacker set.

In this framework, the action space corresponds to the simultaneous selection of beamforming angles for two TXs:

\begin{equation}
A=A_{\mathrm{TX1}}\times A_{\mathrm{TX2}}.
\end{equation}

Each joint action specifies one beam index for each of the two TXs, where each TX beam codebook contains 25 discrete angular values. The RX beam configuration is initialized by uniformly selecting one beam from the corresponding 25-beam codebook for each RX. At every learning iteration, the current RX beams are retained with probability $0.85$ and are replaced by uniformly selected RX beams with probability $0.15$.
For a malicious RX, its nominal beam is overwritten by the corresponding attacker-generated beam when calculating the observed communication performance. Therefore, the RX beams are not selected using the suspicion scores and are not included in the tabular RL action space.

The agent employs an $\varepsilon$-greedy action-selection policy. With probability $1-\varepsilon$, it selects a joint TX action maximizing the Q-value in the current state. When several actions have the same maximum Q-value, the action with the smallest index is selected. With probability $\varepsilon$, the proposed method evaluates every candidate joint TX-beam action $a=(a_1,a_2)$ using the suspicion-weighted utility

\begin{equation}
U^{(t)}(a)
=
\sum_{i=1}^{N_{\mathrm{RX}}}
w_i^{(t)}C_i^{(t)}(a),
\end{equation}

where $C_i^{(t)}(a)$ is the capacity predicted for RX $i$ under
candidate joint TX action $a$ and the current RX beam configuration, and

\begin{equation}
w_i^{(t)}
=
\begin{cases}
1-b_i^{(t)}, & b_i^{(t)}\leq 0.20,\\
-5b_i^{(t)}, & b_i^{(t)}>0.20.
\end{cases}
\end{equation}

The exploratory joint TX action is selected deterministically as

\begin{equation}
a_{\mathrm{explore}}^{(t)}
=
\arg\max_{a\in A}U^{(t)}(a).
\end{equation}

Thus, the proposed exploratory action is not sampled from an
unspecified non-uniform probability distribution. Instead, exploration uses a deterministic suspicion-weighted action proposal. In the plain Q-learning ablation, an exploratory action is selected uniformly from the joint TX action space. The exploration probability is initialized as $\varepsilon=0.3$, multiplied by $0.995$ after each iteration, and lower-bounded by $0.05$.

After selecting the beamforming action, the capacity of each link and the total channel capacity are computed as in \eqref{eq:channelcapacity}. This quantity is then used to construct the reward function for the RL agent.

In this study, the Q-learning algorithm is employed as an off-policy RL method. In this approach, the agent estimates future returns assuming the optimal action is taken, independent of the current policy behavior. The Q-function update rule is given by

\begin{equation}
Q(s,a) \leftarrow Q(s,a)
+ \alpha
\left[
r + \gamma \max_{a'} Q(s',a')
- Q(s,a)
\right].
\end{equation}

where $s$ denotes the current state, $a$ is the action selected in state $s$, $s'$ represents the next state reached after executing action $a$, $a'$ denotes a candidate action in the next state $s'$, $\alpha$ is the learning rate, $\gamma$ is the discount factor, and $r$ is the immediate reward. The term $\max_{a'} Q(s',a')$ represents the greedy estimated optimal future value. This formulation enables convergence toward an optimal greedy policy even when an $\varepsilon$-greedy strategy is used during training. As a result, Q-learning typically exhibits faster convergence toward high-capacity regions, although it may be more sensitive in noisy or adversarial environments.

In contrast, SARSA operates as an on-policy learning algorithm, in which the Q-value update depends on the next action actually selected according to the current policy. Its update rule is given by

\begin{equation}
Q(s,a)
\leftarrow
Q(s,a)
+
\alpha
\left[
r + \gamma Q(s',a')-Q(s,a)
\right].
\end{equation}

where $a'$ denotes the next action selected in state $s'$ according to the current policy. Unlike Q-learning, which uses the maximum next-state action value $\max_{a'}Q(s',a')$, SARSA uses the value $Q(s',a')$ of the action actually selected by the current policy.

For both Q-learning and SARSA, the immediate reward is defined as

\begin{equation}
R
=
R_{\mathrm{efficiency}}
+
R_{\mathrm{drop}}
+
R_{\mathrm{risk}}.
\end{equation}

The capacity-efficiency and capacity-drop components are calculated as

\begin{equation}
R_{\mathrm{efficiency}}
=
\frac{C_{\mathrm{Total}}^{\mathrm{Actual}}}{10^{9}}.
\end{equation}

\begin{equation}
R_{\mathrm{drop}}
=
-5
\left(
\frac{
C_{\mathrm{Total}}
-
C_{\mathrm{Total}}^{\mathrm{Actual}}
}{
C_{\mathrm{Total}}+10^{-12}
}
\right).
\end{equation}

\protect Here, $C_{\mathrm{Total}}$ denotes the aggregate capacity under the intended receiver configuration, whereas
$C_{\mathrm{Total}}^{\mathrm{Actual}}$ denotes the aggregate capacity observed after the attack. Thus, $R_{\mathrm{efficiency}}$ represents the achieved capacity normalized by $10^{9}$, while $R_{\mathrm{drop}}$ penalizes the relative capacity reduction using a weighting factor of $5$.

\protect The security-risk term is initially set to zero. When the belief of the currently most suspicious RX, denoted by $i^{*}$, exceeds $0.25$, the risk penalty is calculated as

\begin{equation}
R_{\mathrm{risk}}
=
-b_{i^{*}}
\frac{C_{i^{*}}}{10^{9}},
\qquad b_{i^{*}}>0.25.
\end{equation}

\protect Finally, the total reward is clipped to the interval $[-1,1]$ to prevent excessively large reward values from producing unstable Q-value updates.

\protect SARSA is selected as a controlled on-policy RL benchmark that complements the off-policy Q-learning method. For a fair comparison, Q-learning and SARSA use the same network realizations, ray-tracing channels, state representation, action space, reward formulation, Bayesian evidence mechanism, and evaluation metrics. Therefore, their
performance differences can be attributed to their respective learning and action-selection policies. A random policy is additionally included to quantify the performance gain obtained from learning and Bayesian-guided decision-making. Exhaustive beam search is used only for the smallest no-attacker configuration as an offline optimization reference, since its computational complexity becomes prohibitive as the number of network nodes increases.

\begin{breakablealgorithm}
\caption{Evidence-Guided Cooperative RL Beamforming for Wireless Adversarial User Detection}
\label{alg:bayesian_rl}
\begin{algorithmic}[1]

\State \textbf{Input:} Network topology, ray-tracing environment,
spatially correlated shadow map, discrete beam codebook $A$,
and RL parameters

\State Initialize the network with
$N_{\mathrm{TX}}=2$ and
$N_{\mathrm{RX}}\in\{4,6,8\}$

\State Generate the TX and RX locations within the considered
deployment area

\State Extract the spatially correlated shadow-fading values
for the TX and RX locations

\State Perform ray tracing for all TX--RX links

\State Store the path loss $PL_k$, propagation phase $\phi_k$,
AoD, and AoA of all propagation rays

\State Associate each RX with a TX using the game-theoretic
association mechanism

\State Define the discrete beam codebook as
\[
A=\{-60^\circ,-55^\circ,\ldots,55^\circ,60^\circ\}
\]

\State Define the cooperative TX action space as
\[
A=A_{\mathrm{TX1}}\times A_{\mathrm{TX2}}
\]

\For{each Monte Carlo realization}

    \State Randomly select
    $N_{\mathrm{Attacker}}\in\{1,2\}$
    malicious RX nodes

    \State Initialize the Bayesian belief vector as
    \[
    \mathbf{b}
    =
    [b_1,b_2,\ldots,b_{N_{\mathrm{RX}}}]^{T}
    \]

    \State Initialize the belief values

    \State Initialize the Q-table

    \State Initialize the exploration probability $\varepsilon$

    \For{each learning iteration}

        \State Identify the RX with the largest belief value:
        \[
        i^{*}
        =
        \arg\max_i b_i
        \]

        \State Map the belief value of RX $i^{*}$ to the
        Bayesian confidence level $L$

        \State Construct the RL state as
        \[
        s=(i^{*},L)
        \]

        \If{Q-learning is used}

            \State Select a joint TX beamforming action $a$
            using the Bayesian-guided $\varepsilon$-greedy policy

        \ElsIf{SARSA is used}

            \State Select a joint TX beamforming action $a$
            using the Bayesian-guided $\varepsilon$-greedy policy

        \EndIf

        \State Apply the selected cooperative TX beamforming action

        \State Determine the corresponding RX beamforming angles

        \State Compute the effective channel gain as
        \[
        G
        =
        \left|
        \mathbf{w}_{\mathrm{r}}^{H}
        \mathbf{H}
        \mathbf{w}_{\mathrm{t}}
        \right|^{2}
        \]

        \State Compute the SINR of each RX:
        \[
        \mathrm{SINR}_j
        =
        \frac{P_{\mathrm{signal}}}
        {P_{\mathrm{interference}}+P_{\mathrm{noise}}}
        \]

        \State Compute the capacity of each RX:
        \[
        C_j
        =
        B\log_2
        \left(
        1+\mathrm{SINR}_j
        \right)
        \]

        \State Compute the total channel capacity:
        \[
        C_{\mathrm{Total}}
        =
        \sum_{j=1}^{N_{\mathrm{RX}}}C_j
        \]

        \State Observe the communication performance of each RX

        \State Update the Bayesian belief values according to
        the observed communication outcomes

        \State Identify the next most probable malicious RX:
        \[
        i^{*\prime}
        =
        \arg\max_i b_i
        \]

        \State Determine the next Bayesian confidence level $L'$

        \State Construct the next state as
        \[
        s'=(i^{*\prime},L')
        \]

        \State Compute the multi-objective reward:
        \[
        R
        =
        R_{\mathrm{efficiency}}
        +
        R_{\mathrm{drop}}
        +
        R_{\mathrm{risk}}
        \]

        \If{Q-learning is used}

            \State Update the Q-value according to
            \[
            Q(s,a)
            \leftarrow
            Q(s,a)
            +
            \alpha
            \left[
            R
            +
            \gamma
            \max_{a'}Q(s',a')
            -
            Q(s,a)
            \right]
            \]

        \ElsIf{SARSA is used}

            \State Select the next action $a'$ using the
            $\varepsilon$-greedy policy

            \State Update the Q-value according to
            \[
            Q(s,a)
            \leftarrow
            Q(s,a)
            +
            \alpha
            \left[
            R
            +
            \gamma Q(s',a')
            -
            Q(s,a)
            \right]
            \]

        \EndIf

        \State Set
        \[
        s\leftarrow s'
        \]

        \State Gradually decrease $\varepsilon$

        \State Record the total channel capacity,
        belief values, and selected beamforming angles

    \EndFor

    \State Estimate the attacker set as the
    $N_{\mathrm{Attacker}}$ RXs with the largest final
    belief values

    \State Compute the attacker-detection performance

    \State Record the best cooperative TX beamforming angles
    and the corresponding RX beamforming angles

    \State Record the total channel capacity and execution time

\EndFor

\State Compute the mean performance over all Monte Carlo realizations

\State \textbf{Output:} Best cooperative TX beamforming angles,
corresponding RX beamforming angles, mean total channel capacity,
attacker-detection performance, execution time, and final belief values

\end{algorithmic}
\end{breakablealgorithm}

The hyperparameters of the proposed framework were determined empirically through preliminary simulation experiments to balance learning stability, convergence speed, and exploration. Specifically, the game-theoretic user association uses a load-balancing coefficient of $\beta=0.8$ and a Softmax temperature of $\tau=0.7$, providing an effective trade-off between utility maximization and association diversity. For the RL algorithms, the initial exploration probability was set to $\epsilon=0.3$ and gradually decayed to a minimum value of $0.05$. The learning rate and discount factor were chosen as $\alpha=0.2$ and $\gamma=0.9$, respectively, to ensure stable learning while maintaining sufficient adaptability to the dynamic wireless environment. Furthermore, the Bayesian evidence accumulation factor was set to $\rho=0.85$ to smooth belief updates while preserving responsiveness to newly observed evidence. These hyperparameter values were kept fixed for all simulation scenarios.

\section{Simulation Results and Discussions} \label{sec:SimulationandDiscussions}

The complete simulation framework, including network generation, ray tracing, game-theoretic RX association, Bayesian belief updating, Q-learning, SARSA, and performance evaluation, was implemented in MATLAB R2026a. All simulations were executed on a computer equipped with a 12th-generation Intel Core i7-12700H processor and 32 GB of RAM.  All execution times reported in this section were measured on the same computational platform.

First, we implement the 3GPP-based system model and evaluate its performance in the presence of interference but without any attacker in the network. The results show that the maximum achievable channel capacity for the 2-TX and 4-RX configuration is 2.36 Gbps, while it is 8.13 Gbps for the 2-TX and 6-RX case, and 15.64 Gbps for the 2-TX and 8-RX setup. These results correspond to the baseline 3GPP channel model without the ray-tracing-based beamforming optimization. Next, the proposed deterministic ray-tracing channel model together with directional beamforming is considered.

More generally, the number of candidate joint beam configurations is

\begin{equation}
N_{\mathrm{config}}
=
N_{\mathrm{angle}}^{N_{\mathrm{TX}}+N_{\mathrm{RX}}},
\label{eq:exhaustive_complexity}
\end{equation}

which demonstrates the exponential dependence of the exhaustive-search complexity on the total number of TXs and RXs. Exhaustive search is therefore employed only as an offline reference for the smallest no-attacker configuration and is not executed during the online operation of the proposed learning framework.

Next, an exhaustive search is performed under beamforming conditions to determine the best antenna steering angles that maximize the total channel capacity. For the baseline network configuration consisting of two TXs and four RXs, the exhaustive search evaluates all possible beamforming combinations and identifies the best combination. The maximum total channel capacity is approximately $2.41$ Gbps, obtained with beam steering angles of $5^\circ$ and $55^\circ$ for TX1 and TX2, respectively. The corresponding best beam steering angles for RX1 to RX4 are $-35^\circ$, $+35^\circ$, $-35^\circ$, and $-35^\circ$. These beamforming angles are subsequently used as the reference solution for evaluating the proposed learning-based framework. As the number of RXs increases, however, the search space grows exponentially. For the scenarios with two TXs and six RXs, the exhaustive search requires evaluating $25^{8}$ possible beamforming combinations, while the search space further increases to $25^{10}$ combinations for the scenario with two TXs and eight RXs.

In the next step, we investigate the performance of the Bayesian Belief method and Q-learning in detecting one or two attackers and determining the best beamforming angles in the presence of attackers. The obtained results are compared with those of the SARSA algorithm and a random baseline, in which the attacker nodes are randomly selected.

\begin{figure}[t]
	\centering
\includegraphics[width=1.05\columnwidth, keepaspectratio]{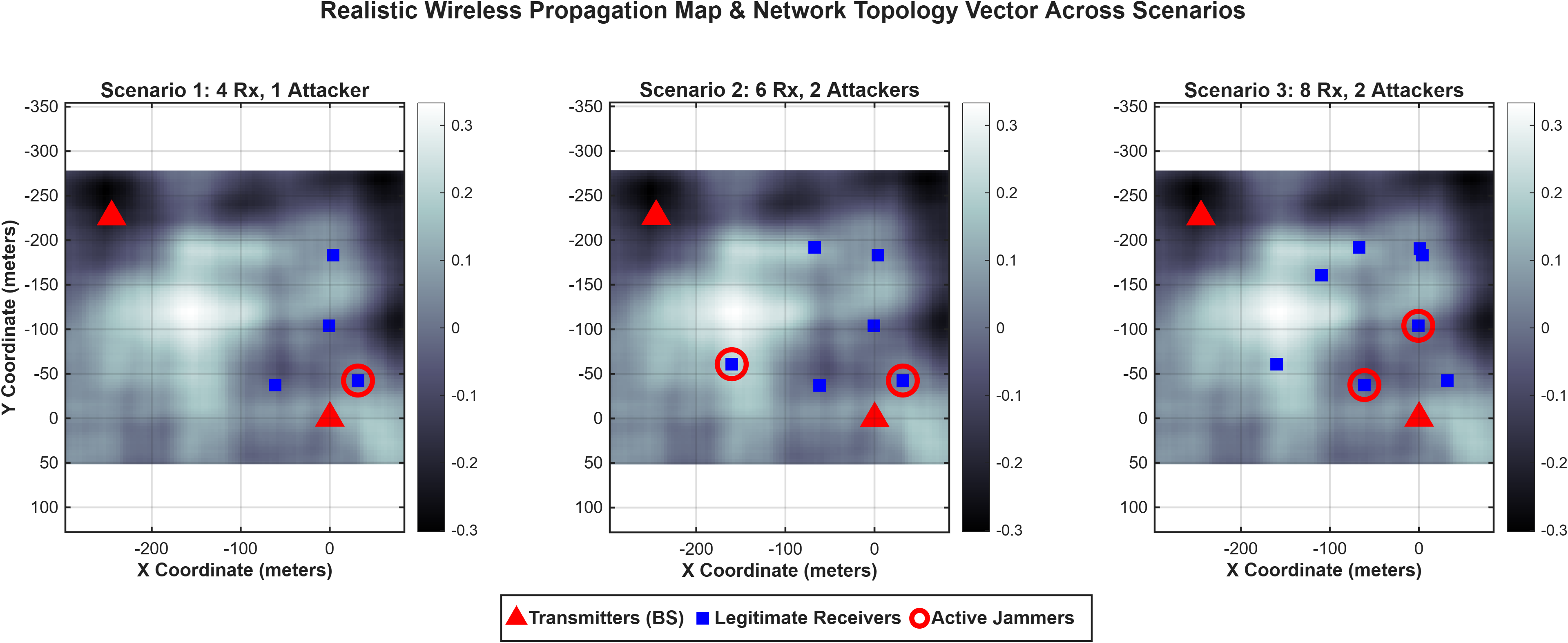}
    \caption{Network topologies with different RX densities and randomly deployed attacker nodes.}
    \label{fig:network_topologies}
\end{figure}

Figure~\ref{fig:network_topologies} illustrates the spatial distribution of TX, legitimate RX, and attacker nodes for the three evaluated scenarios. The left subfigure corresponds to a scenario with $N_{\mathrm{RX}}=4$, consisting of 3 legitimate RXs and 1 attacker. The middle subfigure corresponds to a scenario with $N_{\mathrm{RX}}=6$, consisting of 4 legitimate RXs and 2 attackers. The right subfigure corresponds to a scenario with $N_{\mathrm{RX}}=8$, consisting of 6 legitimate RXs and 2 attackers. To evaluate the robustness of the proposed framework, the attacker identities are randomly selected in each simulation run.

\begin{figure}[t]
	\centering
\includegraphics[width=1\columnwidth, keepaspectratio]{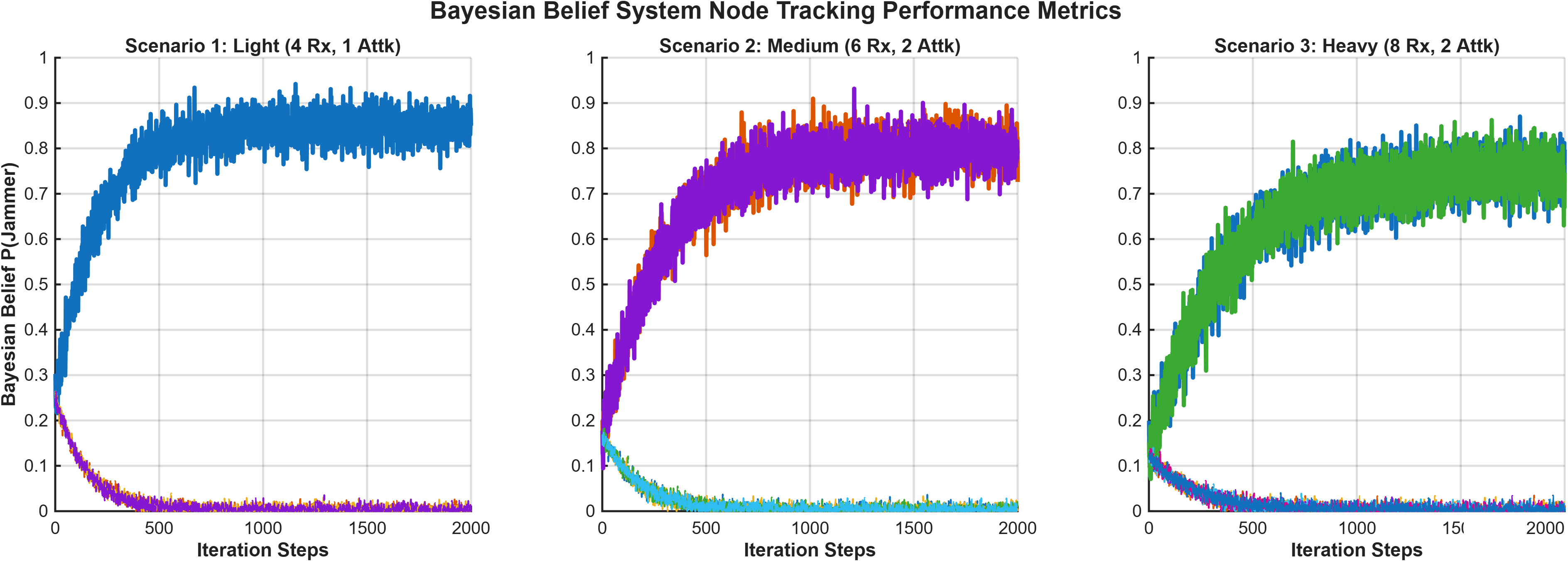}
    \caption{\protect Bayesian belief evolution for attacker-node detection under different network densities. Each panel shows the iteration-wise belief trajectories obtained from one representative realization of the corresponding scenario.}
    \label{fig:bayesian_belief}
\end{figure}

Figure~\ref{fig:bayesian_belief} presents the results obtained from the Bayesian Belief system simulation for detecting and tracking malicious nodes (attackers) across three scenarios with varying levels of complexity. These results demonstrate the effectiveness, robustness, and scalability of the proposed approach. In this figure, the vertical axis represents the posterior belief that each node is malicious, denoted as $P(\text{Attacker})$, while the horizontal axis indicates the number of iterations. As illustrated by Figure~\ref{fig:bayesian_belief}, Algorithm~\ref{alg:bayesian_rl} consistently converges toward a clear separation between malicious and legitimate nodes in all scenarios. Specifically, the belief values associated with attacker nodes gradually increase and stabilize at least above approximately 0.7, reaching values close to 0.9 in some cases, particularly in the low-complexity scenario. In contrast, the belief assigned to legitimate nodes rapidly decreases and remains close to zero, typically below 0.05. This behavior confirms that the Bayesian update mechanism, driven by the discrepancy between predicted and observed channel capacity, effectively enhances the discrimination between normal and abnormal network behavior.

A comparison across scenarios further reveals that convergence is faster and smoother in the first case (3 RXs and 1 attacker), which can be attributed to lower interference levels and a reduced state space. Under these conditions, the observations are less noisy, allowing the belief update process to stabilize more quickly. However, in the second and third scenarios, consisting respectively of 4 and 6 legitimate RXs together with 2 attackers, the increased network density introduces higher levels of interference and observation
uncertainty. This leads to slower initial convergence and slightly higher variance in the steady-state region. Despite the increased complexity, the algorithm remains capable of correctly identifying malicious nodes with satisfactory accuracy. Even in the most challenging scenario, the belief values for attacker nodes remain stable (approximately 0.65 to 0.85) without exhibiting instability or divergence. This indicates that the proposed Bayesian framework maintains its robustness as the network size increases. 

These results indicate that the proposed Bayesian framework consistently maintains reliable attacker discrimination across all considered network scenarios. Despite the increased interference and larger state space in denser deployments, the posterior beliefs converge to stable values that clearly distinguish malicious nodes from legitimate users. This behavior demonstrates the robustness of the Bayesian inference mechanism and its ability to maintain stable detection performance as network complexity increases.

\begin{figure}[t]
	\centering
\includegraphics[width=0.75\columnwidth, keepaspectratio]{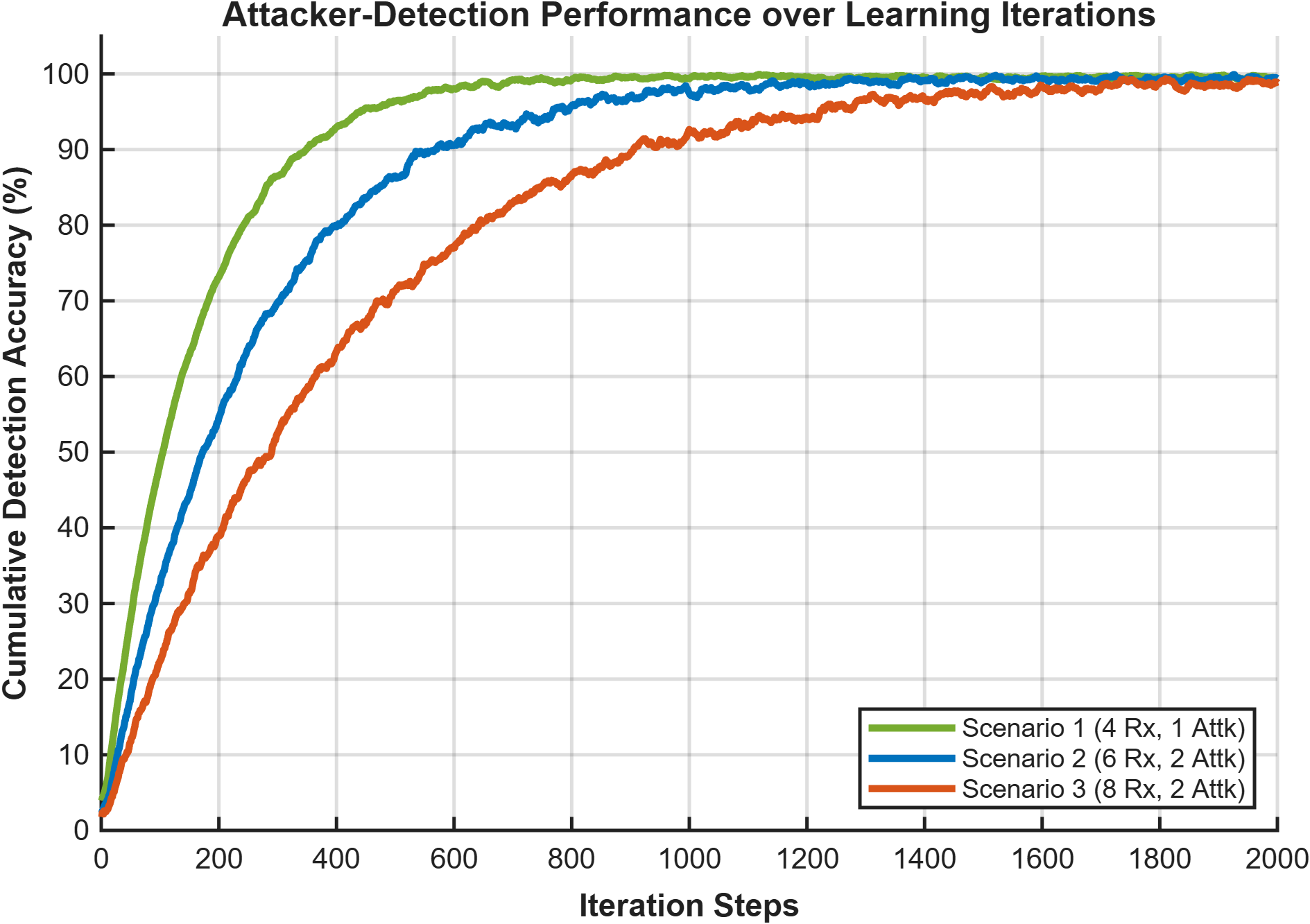}
\caption{\protect Cumulative attacker-detection percentage over the learning iterations for three network scenarios. Each curve corresponds to one representative 2000-iteration run and shows the detection performance accumulated up to each iteration.}
    \label{fig:accuracy}
\end{figure}

Moreover, Figure~\ref{fig:accuracy} illustrates the attacker-detection performance of the proposed framework over the learning iterations, highlighting its convergence behavior and effectiveness in identifying malicious RX nodes. In this figure, the vertical axis represents the attacker detection performance (in percentage), while the horizontal axis corresponds to simulation iterations. The curves corresponding to all three scenarios demonstrate stable learning behavior under different network densities, with the attacker-detection performance progressively improving over the learning iterations and eventually approaching 100\%. This trend indicates the effectiveness of the proposed framework for the considered simulation scenarios.

A comparative analysis of the scenarios shows that the convergence speed depends on both the total number of RXs and the number of attackers. In Scenario~1, with $N_{\mathrm{RX}}=4$ and one attacker, the smaller network size and lower level of interference result in the fastest convergence, with the attacker-detection performance approaching $100\%$ before approximately 500 iterations. Scenario~2, with $N_{\mathrm{RX}}=6$ and two attackers, converges more gradually and reaches a stable high attacker-detection performance around iteration 1000. Scenario~3, with $N_{\mathrm{RX}}=8$ and two attackers, has the same number of attackers as Scenario~2 but a larger network and more legitimate RXs. Therefore, its slower initial convergence should be interpreted as an empirical consequence of the larger and more interference-prone network realization rather than a larger attacker fraction. Nevertheless, the learning process exhibits stable convergence throughout the simulation. After approximately 1500 iterations, the attacker-detection performance continues to improve, eventually exceeding $98\%$. These results demonstrate the effectiveness of the proposed framework for the considered network configurations.

Regarding the relationship between Figures ~\ref{fig:bayesian_belief} and \ref{fig:accuracy}, the main distinction lies in the perspective of evaluation. Figure~\ref{fig:bayesian_belief} provides an internal, probabilistic view of the system behavior, showing the Bayesian belief evolution for each node (i.e., how the system updates its confidence about whether a node is malicious or normal at each iteration). This representation naturally exhibits fluctuations due to noisy observations and continuous probabilistic updates. In contrast, Figure~\ref{fig:accuracy} presents a higher-level, aggregated performance metric, combining the overall decision outcome (based on Bayesian inference and Q-learning) into a single accuracy measure. While Figure~\ref{fig:bayesian_belief} reflects the internal decision dynamics and uncertainty of the model, Figure~\ref{fig:accuracy} demonstrates the final external performance, showing that despite internal fluctuations, the system consistently achieves highly stable and near-optimal classification accuracy. Therefore, the gradual convergence of the Bayesian posterior beliefs shown in Figure~\ref{fig:bayesian_belief} directly translates into the increasing detection accuracy observed in Figure~\ref{fig:accuracy}.

\begin{figure}[t]
    \centering
    \includegraphics[width=\columnwidth,keepaspectratio]{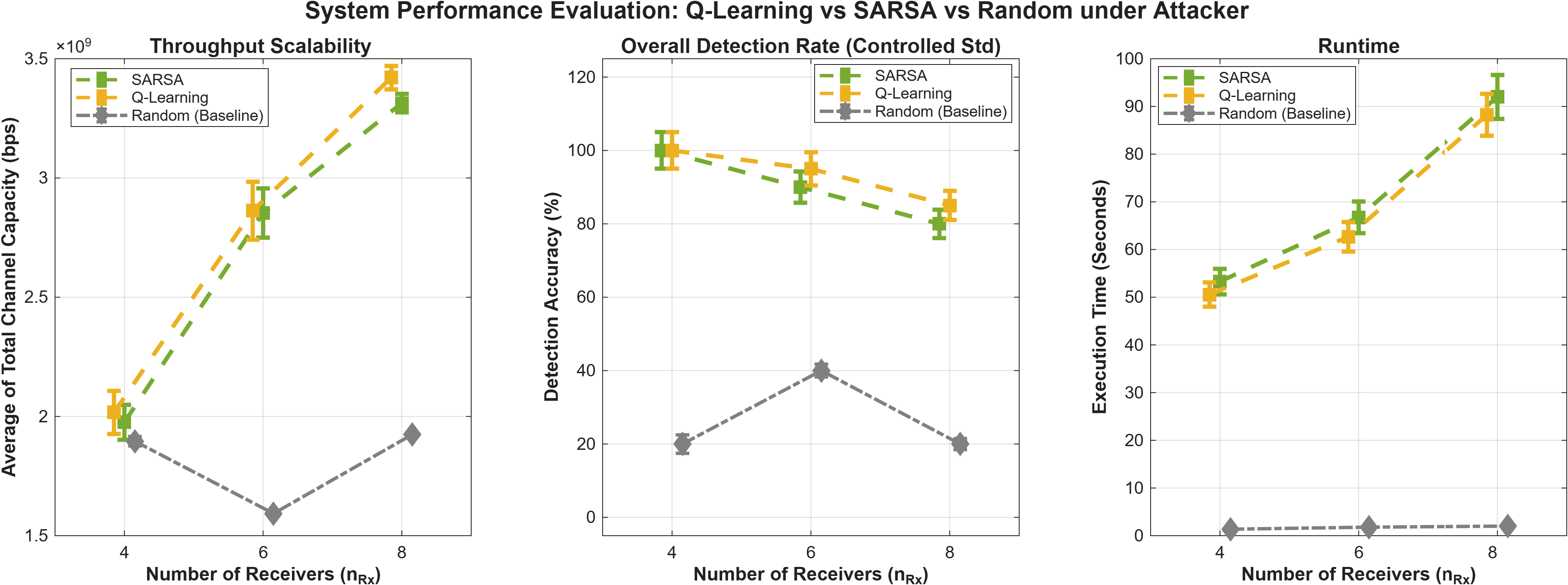}
    \caption{\protect Comparison of Q-learning, SARSA, and the random policy in terms of aggregate channel capacity, attacker-detection accuracy, and execution time for networks with 4, 6, and 8 RXs.
    Bars represent the means over 20 independent Monte Carlo realizations, and error bars indicate the corresponding standard deviations.}
    \label{fig:rl_comparison}
\end{figure}

Next, Figure~\ref{fig:rl_comparison} compares the performance of the proposed RL framework under adversarial conditions using three decision-making strategies, namely Q-learning, SARSA, and a random baseline. The reported results are averaged over 20 independent Monte Carlo realizations, while the error bars represent the corresponding standard deviations, illustrating the consistency of each method across different network realizations. The comparison is conducted in terms of throughput, attacker detection accuracy, and computational complexity for three network sizes, namely $N_{\mathrm{RX}}\in\{4,6,8\}$. As shown in the left subplot, both RL algorithms achieve progressively higher throughput as the number of RXs increases. This behavior is primarily attributed to the larger number of simultaneously served communication links together with the ability of the learning-based beamforming strategy to efficiently exploit the additional spatial degrees of freedom, thereby increasing the overall network capacity despite the stronger interference in denser deployments. Among the evaluated methods, Q-learning consistently achieves the highest throughput, reaching approximately $2.0\times10^{9}$, $2.86\times10^{9}$, and $3.42\times10^{9}$ bps for $N_{\mathrm{RX}}=4$, $6$, and $8$, respectively. SARSA follows a similar trend with slightly lower average capacities of approximately $1.97\times10^{9}$, $2.85\times10^{9}$, and $3.31\times10^{9}$ bps. In contrast, the random baseline provides considerably lower throughput, fluctuating around $1.9\times10^{9}$ bps without exhibiting a consistent scalability trend. The relatively small standard deviations indicate that both learning-based methods maintain stable throughput performance across different Monte Carlo realizations. Since $C_{\mathrm{Total}}$ represents the sum capacity over all RXs, the observed increase with $N_{\mathrm{RX}}$ should be interpreted as an increase in aggregate network throughput rather than necessarily an improvement in the average throughput achieved by each individual RX.

The middle subplot presents the attacker detection performance. Both Q-learning and SARSA successfully detect nearly all attackers in the smallest network configuration, achieving detection accuracies close to $100\%$. As the number of RXs increases, the detection task becomes more challenging because of the enlarged state-action space and the increased uncertainty introduced by simultaneous transmissions. Nevertheless, Q-learning consistently maintains higher detection accuracy than SARSA, achieving approximately $95\%$ and $84\%$ for $N_{\mathrm{RX}}=6$ and $8$, respectively, whereas SARSA achieves approximately $90\%$ and $80\%$. By comparison, the random policy exhibits significantly poorer performance, providing detection accuracies of only about $20\%$, $40\%$, and $20\%$ for the three network sizes. These results demonstrate that RL substantially improves the reliability of malicious-user identification compared with random decision making while maintaining relatively low performance variability.

The computational complexity results, illustrated in the right subplot of Figure~\ref{fig:rl_comparison}, show that the execution time of both RL algorithms increases with network size because of the growth of the state-action space and the additional Q-value update operations required during learning. Q-learning requires approximately $50$, $62$, and $88$ seconds for $N_{\mathrm{RX}}=4$, $6$, and $8$, respectively, while SARSA requires approximately $53$, $66$, and $92$ seconds. In contrast, the random baseline performs neither learning nor policy optimization and therefore maintains an execution time of only about $1$-$2$ seconds for all scenarios. Although RL introduces additional computational overhead, the runtime increases moderately over the considered network sizes and is justified by the considerable improvements achieved in both throughput and attacker detection. Moreover, Q-learning consistently requires slightly less execution time than SARSA while simultaneously providing superior communication performance.

Table~\ref{tab:optimal_beams} presents the best-found beamforming angles selected by the Q-learning and SARSA algorithms for the considered network configurations. For each scenario, the table reports the optimal beam directions selected for the two cooperative TXs together with the corresponding steering angles of all RXs. The receive angles associated with the detected attacker nodes are underlined for clarity. As the number of RXs increases, the selected beam directions change to accommodate the different network topology and propagation conditions. While both algorithms modify their beam configurations according to the communication environment, the solutions obtained by Q-learning generally lead to higher throughput and more reliable attacker detection than those produced by SARSA.

\begin{table*}[t]
\centering
\caption{Joint TX beam angles selected by Q-learning and SARSA and the corresponding RX beam angles.}
\label{tab:optimal_beams}
\renewcommand{\arraystretch}{1.25}
\begin{tabular}{ccccc}
\toprule
\textbf{$N_{\mathrm{RX}}$} &
\textbf{Algorithm} &
\textbf{TX$_1$} &
\textbf{TX$_2$} &
\textbf{Corresponding RX Beam Angles ($^\circ$)}\\
\midrule

\multirow{2}{*}{4}
& Q-learning
& 60 & -40
& $60,\,-45,\,\underline{-40},\,-10$ \\

& SARSA
& 60 & 50
& $60,\,-60,\,\underline{5},\,-30$ \\

\midrule

\multirow{2}{*}{6}
& Q-learning
& 60 & -60
& $60,\,35,\,40,\,-10,\,\underline{5},\,\underline{60}$ \\

& SARSA
& 60 & -60
& $60,\,60,\,-25,\,-55,\,\underline{45},\,\underline{15}$ \\

\midrule

\multirow{2}{*}{8}
& Q-learning
& 60 & -60
& $60,\,-55,\,\underline{-45},\,-20,\,50,\,-5,\,-25,\,\underline{10}$ \\

& SARSA
& 60 & -60
& $-60,\,-60,\,\underline{-30},\,-25,\,50,\,-50,\,-30,\,\underline{-45}$ \\

\bottomrule
\end{tabular}
\end{table*}

These results indicate that Q-learning offers the most favorable balance between throughput, attacker detection accuracy, and computational complexity among the evaluated methods. Although both RL algorithms substantially outperform the random baseline, Q-learning consistently provides higher throughput, better detection performance, and slightly lower computational cost. These observations show that, under the evaluated network configurations and attacker settings, incorporating RL into the beam-selection process improves aggregate communication performance and attacker-detection accuracy compared with the considered random policy.

\begin{figure}[t]
    \centering
    \includegraphics[width=\linewidth]
    {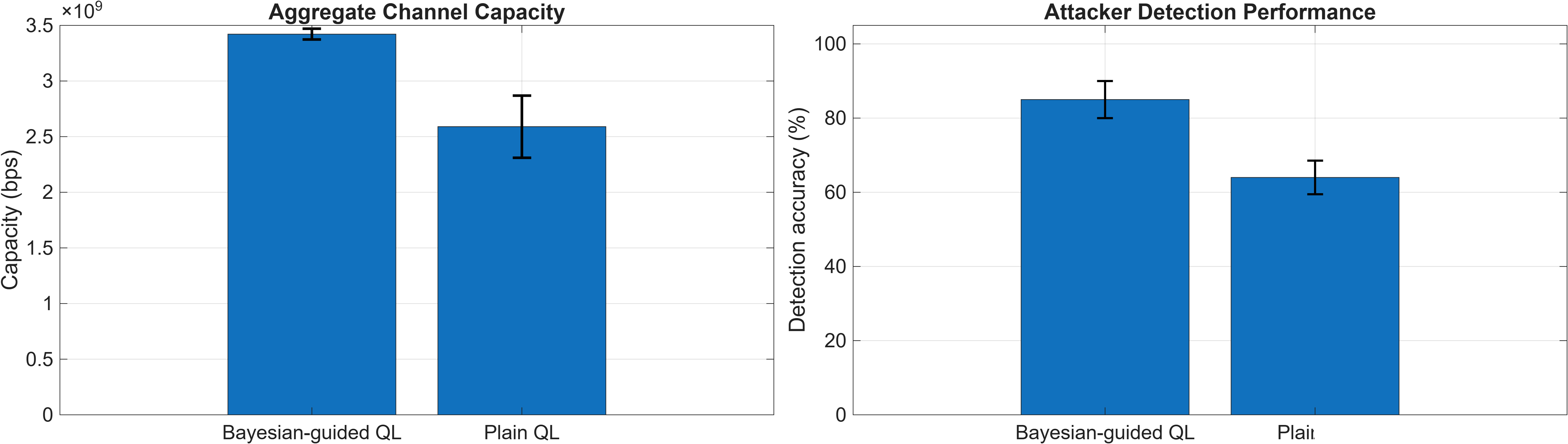}
    \caption{\protect Ablation comparison of the proposed
    Bayesian-guided Q-learning method and conventional plain Q-learning with uniform $\epsilon$-greedy exploration for the 8-RX/2-attacker scenario. Bars and error bars represent the mean and one standard deviation over 20 independent realizations, respectively. Both methods were evaluated using identical network realizations and common RL hyperparameters.}
    \label{fig:bayesian_guided_ablation}
\end{figure}

To isolate the contribution of Bayesian guidance, the proposed Bayesian-guided Q-learning method was compared with a conventional plain Q-learning baseline using uniform $\epsilon$-greedy exploration for the 8-RX/2-attacker scenario. Both methods were evaluated using the same network realizations, channel models, beamforming action space, learning
rate, discount factor, and $\epsilon$ schedule. The plain Q-learning baseline employed a conventional Q-table to learn the joint TX-beam actions and identify malicious RXs directly from the received reward feedback. Actions were selected using standard uniform $\epsilon$-greedy exploration, without receiver-specific Bayesian belief updates, belief-guided states, or suspicion-guided action selection. Therefore, the reported attacker-detection accuracy was obtained directly from the learned Q-values rather than from a separate Bayesian detector.
As shown in Fig.~ \ref{fig:bayesian_guided_ablation}, the Bayesian-guided Q-learning method achieved an average aggregate channel capacity of $3.42 \pm 0.05$~ Gbps, compared with $2.59 \pm 0.28$~ Gbps for plain Q-learning, corresponding to an improvement of approximately $32\%$. Moreover, the Bayesian-guided Q-learning method achieved an attacker-detection accuracy of $84\%$, whereas plain Q-learning
achieved $64\%$, representing an improvement of 20 percentage points. The capacity values represent the mean and standard deviation over 20 independent realizations. These results demonstrate that the performance gains are not solely attributable to the use of conventional Q-learning. Instead, incorporating receiver-specific Bayesian belief information into the learning and action-selection process results in
higher aggregate channel capacity and attacker-detection accuracy under the evaluated scenario.

In addition, for the Bayesian-guided Q-learning method in the
8-RX/2-attacker scenario, the aggregate channel capacity remained within a relatively narrow range under the considered hyperparameter variations.
Specifically, the achieved capacities were approximately
$3.29\times10^{9}$, $3.42\times10^{9}$, and $3.35\times10^{9}$~bps for $\alpha=0.1$, $0.2$, and $0.3$, respectively, with $\gamma=0.9$. For a fixed learning rate of $\alpha=0.2$, capacities of approximately $3.23\times10^{9}$, $3.42\times10^{9}$, and $3.39\times10^{9}$~bps were obtained for $\gamma=0.8$, $0.9$, and $0.95$, respectively. These results
indicate that the proposed method maintains a consistently high aggregate channel capacity across the evaluated learning-rate and discount-factor settings and does not depend on a single narrowly fine-tuned selection of $\alpha$ and $\gamma$.

\section{Conclusion}\label{sec:conclusions}

In this study, a hybrid intelligent framework combining RL and Bayesian inference was developed for secure and adaptive beamforming in mmWave networks operating under dynamic and adversarial conditions. The objective of the proposed framework was to enhance beam steering decisions, improve aggregate network capacity, mitigate attacker-induced performance degradation, and identify malicious nodes in a multi-RX wireless environment. In the initial stage, a 3GPP-inspired system model was implemented and evaluated under a no-attacker scenario. The results indicated that the aggregate channel capacity increases with the number of RXs in the considered configurations, as expected from the summation over a larger number of links. An exhaustive search was then conducted for the baseline configuration to identify reference beamforming configurations. The framework was subsequently evaluated in the presence of attackers and under different network scalability settings. Based on 20 independent Monte Carlo simulations, the RL-based methods outperform the random baseline in terms of aggregate channel capacity and attacker detection performance in the considered scenarios, at the cost of higher runtime. Among the evaluated methods, Q-learning achieves the highest total channel capacity and attacker detection accuracy while requiring slightly less runtime than SARSA. The beamforming results show that both Q-learning and SARSA identify stable beam configurations, although the selected angles need not coincide because the two algorithms update their action values under different policies. The results support the effectiveness of the proposed Bayesian-guided RL framework for joint beamforming and attacker-aware decision-making in the considered mmWave network scenarios.
Despite the promising results, the evaluation is limited to simulated urban scenarios with fixed node positions and a known number of attackers.
Future work will extend the framework toward a unified learning-based model that jointly addresses attacker detection, beamforming, and dynamic power allocation. Deep RL methods will also be investigated to support larger state spaces and continuous actions, potentially improving capacity, robustness, and energy efficiency.

\vspace{6pt}

\authorcontributions{
All authors participated in the conceptualization, methodology, mathematical analysis, and project management. PG participated in writing the original draft, and OG  participated in reviewing and editing the original draft. 
}
{
\funding{The work of OG was partially supported by the ZENITH Research and Leadership Career Development Fund under Grant ID23.01, EU COST Action 6G-PHYSEC,  Swedish Foundation for Strategic Research (SSF) under Grant ID24-0087, and German Federal Ministry of Research, Technology and Space (BMFTR) 6GEM+ Research Hub under the Grants 16KIS2412 and 16KISS005.}
}
\institutionalreview{Not applicable.}

\informedconsent{Not applicable.}

\dataavailability{Not applicable.} 

\acknowledgments{During the preparation of this manuscript, the authors used OpenAI ChatGPT 5.5 for the purposes of assisting with minor language editing of the authors' own text. The authors have reviewed and edited the output and take full responsibility for the content of this publication.}

\conflictsofinterest{The authors declare no conflicts of interest.} 

\abbreviations{Abbreviations}{
The following abbreviations are used in this manuscript:
\\

\noindent 
\begin{tabular}{@{}ll}
3GPP & 3rd Generation Partnership Project\\
5G NR & 5th Generation New Radio\\
6G & 6th Generation (networks)\\
AoA & Angle of Arrival\\
AoD & Angle of Departure\\
BS & Base Station(s)\\
CIR & Channel Impulse Response\\
DRL & Deep Reinforcement Learning\\
FR2 & Frequency Range 2\\
ISAC & Integrated Sensing and Communication\\
ISD & Inter-Site Distance\\
IRS & Intelligent Reflective Surface(s)\\
IoT & Internet of Things\\
LOS & Line of Sight\\
ML & Machine Learning\\
mmWave & millimeter Wave\\
MTD & Moving Target Defense\\
NF\_UE & Noise Figure of User Equipment\\
NLOS & Non-Line of Sight\\
NOMA & Non-Orthogonal Multiple Access\\
OSM & OpenStreetMap\\
PPP & Poisson Point Process\\
QoS & Quality of Service\\
RL & Reinforcement Learning\\
RX & Receiver\\
SARSA & State–Action–Reward–State–Action\\
SBR & Shooting and Bouncing Rays\\
SINR & Signal-to-Interference-plus-Noise Ratio\\
TR 38.901 & Technical Report 38.901 (3GPP channel model standard)\\
TX & Transmitter\\
UAV & Unmanned Aerial Vehicle\\
UE & User Equipment\\
UHF & Ultra High Frequency\\
UMi & Urban Micro-cellular (scenario)\\
\end{tabular}
}

\begin{adjustwidth}{-\extralength}{0cm}
\reftitle{References}

\bibliography{referencesEntropy}

@ARTICLE{Ding2022DigitalTwin,
  author={Ding, C. and Wang, I. and Ho, H.},
  journal={IEEE Trans. Green Commun. Netw.},
  title={Digital-twin-enabled city-model-aware deep learning for dynamic channel estimation in urban vehicular environments},
  year={2022},
  volume={6},
  number={3},
  pages={1604--1612}
}

@ARTICLE{Geok2018RayTracing,
  author={Geok, T. K. and Hossain, F. and Liew, C. P.},
  journal={Int. J. Commun. Antenna Propag. (IRECAP)},
  title={A comprehensive review of efficient ray-tracing techniques for wireless communication},
  year={2018},
  volume={8},
  number={2},
  \\doi={10.15866/irecap.v8i2.13797}
}

@ARTICLE{Sheen2020DigitalTwinRIS,
  author={Sheen, B. and Yang, J. and Feng, X. and Chowdhury, M.},
  title={A digital twin for reconfigurable intelligent surface assisted wireless communication},
  journal={arXiv},
  volume={abs/2009.00454},
  year={2020},
  \\doi={10.48550/arXiv.2009.00454}
}

@INPROCEEDINGS{Jalali2023SWIPT,
  author={Jalali, J. and Khalili, A. and Rezaei, A. and Famaey, J. and Saad, W.},
  booktitle={IEEE 20th Consumer Communications \& Networking Conference (CCNC)},
  title={Power-efficient antenna switching and beamforming design for multi-user SWIPT with non-linear energy harvesting},
  year={2023},
  \\doi={10.1109/CCNC51644.2023.10059879}
}

@ARTICLE{Bakhtiar2023Beamforming,
  author={Bakhtiar, A. K. and Haitham, K. A.},
  journal={Digital Signal Processing},
  title={A novel beamforming technique using MmWave antenna arrays for 5G wireless communication networks},
  year={2023},
  volume={136},
  pages={103917},
  \\doi={10.1016/j.dsp.2023.103917}
}

@ARTICLE{Chary2024AIBeamforming,
  author={Chary, M. K. and Krishna, C. H. V. and Krishna, D. R.},
  journal={AEU - Int. J. Electron. Commun.},
  title={Accurate channel estimation and hybrid beamforming using artificial intelligence for massive MIMO 5G systems},
  year={2024},
  \\doi={10.1016/j.aeue.2023.154971}
}

@INCOLLECTION{Geranmayeh2022BeamSelection,
  author={Geranmayeh, P. and Sedunova, E. and Grass, E.},
  booktitle={Recent Advances in Communication Networks and Embedded Systems (ICCNT)},
  title={Efficient beam selection for increased overall wireless network capacity},
  year={2022},
  publisher={Springer},
  address={Cham, Switzerland},
  \\doi={10.1007/978-3-031-59619-3_2}
}

@ARTICLE{Ihsan2022IRS,
  author={Ihsan, A. and Chen, W. and Asif, M. and Khan, W. U. and Wu, Q. and Li, J.},
  journal={IEEE Trans. Green Commun. Netw.},
  title={Energy-efficient IRS-aided NOMA beamforming for 6G wireless communications},
  year={2022},
  \\doi={10.1109/TGCN.2022.3209617}
}

@INPROCEEDINGS{Jiang2022ICC,
  author={Jiang, W. and Schotten, H. D.},
  booktitle={IEEE International Conference on Communications (ICC)},
  title={Initial access for millimeter-wave and terahertz communications with hybrid beamforming},
  year={2022},
  \\doi={10.1109/ICC45855.2022.9838386}
}

@ARTICLE{Qi2021UAVBeamforming,
  author={Qi, F. and Li, W. and Yu, P. and Feng, L. and Zhou, F.},
  journal={EURASIP J. Wirel. Commun. Netw.},
  title={Deep learning-based backcom multiple beamforming for 6G UAV IoT networks},
  year={2021},
  \\doi={10.1186/s13638-021-01932-4}
}

@ARTICLE{Ji2021RayTracingIndoor,
  author={Ji, Z. and Li, B. H. and Wang, H. X. and Chen, H. Y. and Sarkar, T. K.},
  journal={IEEE Antennas Propag. Mag.},
  title={Efficient ray-tracing methods for propagation prediction for indoor wireless communications},
  year={2001},
  volume={43},
  number={2},
  pages={41--49}
}

@ARTICLE{Hsiao2017RayTracing,
  author={Hsiao, A. Y. and Yang, C. F. and Wang, T. S. and Lin, I. and Liao, W. J.},
  journal={IEEE Int. Symp. Antennas Propag.},
  title={Ray-tracing simulations for millimeter wave propagation in 5G wireless communications},
  year={2017},
  \\doi={10.1109/APUSNCURSINRSM.2017.8072993}
}

@ARTICLE{Yun2015RayTracingTheory,
  author={Yun, Z. and Iskander, M. F.},
  journal={IEEE Access},
  title={Ray-tracing for radio propagation modeling: principles and applications},
  year={2015},
  volume={3},
  pages={1089--1100},
  \\doi={10.1109/ACCESS.2015.2453991}
}

@ARTICLE{Tiberi2009UWB,
  author={Tiberi, G. and Bertini, S. and Malik, W. K. and Monorchio, A. and Edwards, D. J. and Man, G.},
  journal={IEEE Trans. Antennas Propag.},
  title={Analysis of realistic ultrawideband indoor communication channels by using an efficient ray-tracing based method},
  year={2009},
  volume={57},
  number={3},
  pages={777--785}
}

@ARTICLE{Geranmayeh2025DQN,
  author={Geranmayeh, P. and Grass, E.},
  journal={IEEE Access},
  title={Optimization of Beamforming and Transmit Power Using DQN and Comparison With Traditional Techniques},
  year={2025},
  volume={13},
  pages={94275--94285},
  \\doi={10.1109/ACCESS.2025.3573096}
}

@INPROCEEDINGS{Vivekanand2024DRL,
  author={Vivekanand, C. V. and Talele, G. and Jaishanthi, B. and Gopalan, S. H. and Maheswari, K. and Maheswari, B. U.},
  booktitle={9th International Conference on Science, Technology, Engineering and Mathematics (ICONSTEM)},
  title={Deep reinforcement learning for resource allocation in wireless communication systems},
  year={2024},
  \\doi={10.1109/ICONSTEM60960.2024.10568620}
}

@ARTICLE{Lei2018Game,
  author={Lei, C. and Zhang, H. Q. and Wan, L. M. and Liu, L. and Ma, D. H.},
  journal={Comput. Commun.},
  title={Incomplete information Markov game theoretic approach to strategy generation for moving target defense},
  year={2018},
  volume={116},
  pages={184--199}
}

@ARTICLE{Jiang2021Game,
  author={Jiang, L. and Hengwei, Z. and Wang, J.},
  journal={J. Electron.},
  title={Optimal decision-making method of moving target defense based on multi-stage Markov signal game},
  year={2021},
  volume={49},
  number={3},
  pages={527--535}
}

@INPROCEEDINGS{Teofilo2012Poker,
  author={Te{\'o}filo, L. F. and Passos, N. and Reis, L. P. and Cardoso, H. L.},
  booktitle={International Conference on Autonomous and Intelligent Systems},
  title={Adapting strategies to opponent models in incomplete information games: A reinforcement learning approach for poker},
  year={2012},
  pages={220--227}
}

@INPROCEEDINGS{Yu2023Covert,
  author={Yu, F. and Qian, L. and Zhang, H. and Wang, X.},
  booktitle={9th International Conference on Mechanical and Electronics Engineering (ICMEE)},
  title={Covert communication assisted by friendly jammer with jamming-monitoring channel uncertainty},
  year={2023},
  pages={43--48}
}

@ARTICLE{Esmaili2026Covert,
  author={Esmaili, E. and Hajizadeh, R. and Forouzesh, M.},
  journal={Sci. Rep.},
  title={Machine learning based detection of covert communications under jamming interference},
  year={2026},
  volume={16},
  number={53830},
  \\doi={10.1038/s41598-026-53830-8}
}

@ARTICLE{Elsadig2022Covert,
  author={Elsadig, M. A. and Gafar, A.},
  journal={IEEE Access},
  title={Covert channel detection: machine learning approaches},
  year={2022},
  volume={10},
  pages={38391--38405}
}

@ARTICLE{Ren2023UAV,
  author={Ren, Z. and Zhang, D. and Tang, S. and Xiong, W. and Yang, S. H.},
  journal={Defence Technol.},
  title={Cooperative maneuver decision making for multi-UAV air combat based on incomplete information dynamic game},
  year={2023},
  volume={27},
  pages={308--317}
}

@ARTICLE{Yao2024BayesianGame,
  author={Yao, P. and Jiang, Z. and Yan, B. and Yang, Q. and Wang, W.},
  journal={Inf. Sci.},
  title={Bayesian and stochastic game joint approach for cross-layer optimal defensive decision-making in industrial cyber-physical systems},
  year={2024},
  volume={662},
  pages={120216--120251}
}

@ARTICLE{Mamaghani2026ISAC,
  author={Mamaghani, M. T. and Zhou, X. and Yang, N. and Swindlehurst, A. L.},
  journal={IEEE J. Sel. Areas Commun.},
  title={Securing integrated sensing and communication against a mobile adversary: A Stackelberg game with deep reinforcement learning},
  year={2026},
  volume={44},
  number={4},
  pages={942--958},
  \\doi={10.1109/JSAC.2025.3611404}
}

@ARTICLE{Jere2023BayesianJamming,
  author={Jere, Shashank and Wang, Ying and Aryendu, Ishan and Dayekh, Shehadi and Liu, Lingjia},
  title={Bayesian Inference-assisted Machine Learning for Near Real-Time Jamming Detection and Classification in 5G New Radio (NR)},
  journal={arXiv},
  volume={abs/2304.13660},
  year={2023}
}

@INPROCEEDINGS{Huang2022VANET,
  author={Huang, F. and Li, Q. and Zhao, J.},
  booktitle={IEEE/CIC International Conference on Communications in China (ICCC) Workshops},
  title={Trust Management Model of VANETs Based on Machine Learning and Active Detection Technology},
  year={2022},
  pages={412--416},
  \\doi={10.1109/ICCCWorkshops55477.2022.9896700}
}

@article{Geranmayeh2026ISAC,
  author  = {Geranmayeh, Parmida and G{\"u}nl{\"u}, Onur},
  title   = {6G Sensing Security: Distributed Game-Theoretic
             Reinforcement Learning for Urban Beamforming and
             Attacker Detection},
  journal = {arXiv preprint arXiv:2607.06115},
  year    = {2026},
  note    = {Available online: \url{https://arxiv.org/abs/2607.06115}}
}

@inproceedings{Zhou2026FeedbackLunch,
  author    = {Zhou, Yingyao and Devroye, Natasha and G{\"u}nl{\"u}, Onur},
  title     = {Feedback Lunch: Learned Feedback Codes for Secure Communications},
  booktitle = {2026 ACM Workshop on Wireless Security and Machine Learning (WiseML '26)},
  address   = {Saarbr{\"u}cken, Germany},
  publisher = {Association for Computing Machinery},
  year      = {2026},
  pages     = {1--6},
  \\doi       = {10.1145/3811880.3815100}
}

@ARTICLE{OnurRoleofISAC,
  author={Qaisar, Muhammad Umar Farooq and others},
  journal={IEEE Transactions on Network Science and Engineering}, 
  title={The Role of {ISAC in 6G Networks: Enabling} Next-Generation Wireless Systems}, 
  year={2026},
  volume={13},
  number={},
  pages={7825-7861},
  }

@article{OnurSecureISACJournal,
  author={Günlü, Onur and Bloch, Matthieu R. and Schaefer, Rafael F. and Yener, Aylin},
  journal={IEEE Journal on Selected Areas in Information Theory}, 
  title={Secure Integrated Sensing and Communication}, 
  year={2023},
  volume={4},
  number={},
  pages={40-53},
}

@article{Onur2026secureISACTutorial,
  title={Secure Integrated Sensing and Communication: {Information} Theory Offers Insights},
  author={Welling, Truman and G{\"u}nl{\"u}, Onur and Yener, Aylin},
  journal={arXiv preprint arXiv:2605.08106},
  year={2026}
}

@ARTICLE{OnurSecurityTutorial,
  author={Bloch, Matthieu and others},
  journal={IEEE Journal on Selected Areas in Information Theory (JSAIT)}, 
  title={An Overview of Information-Theoretic Security and Privacy: Metrics, Limits and Applications}, 
  year={2021},
  volume={2},
  number={1},
  pages={5-22},
  \\doi={10.1109/JSAIT.2021.3062755}}

@techreport{ETSI_TR_138901_v1610,
  author      = {{ETSI}},
  title       = {{5G; Study on Channel Model for Frequencies from 0.5 to 100 GHz}},
  institution = {{European Telecommunications Standards Institute (ETSI)}},
  type        = {{3GPP TR 38.901}},
  number      = {{Version 16.1.0, Release 16}},
  year        = {2020},
}

@ARTICLE{Gudmundson1991Correlation,
  author={Gudmundson, Mikael},
  journal={Electronics Letters},
  title={Correlation Model for Shadow Fading in Mobile Radio Systems},
  year={1991},
  volume={27},
  number={23},
  pages={2145--2146},
  \\doi={10.1049/el:19911328}
}

@techreport{ITU_R_M2135_1,
  author      = {{ITU-R}},
  title       = {Guidelines for Evaluation of Radio Interface Technologies for {IMT-Advanced}},
  number      = {M.2135-1},
  address     = {Geneva, Switzerland},
  year        = {2009},
}

\section*{Short Biography of Authors}

\bio{%
\raisebox{-0.35cm}{%
\includegraphics[width=3.5cm,height=5.3cm,clip,keepaspectratio]{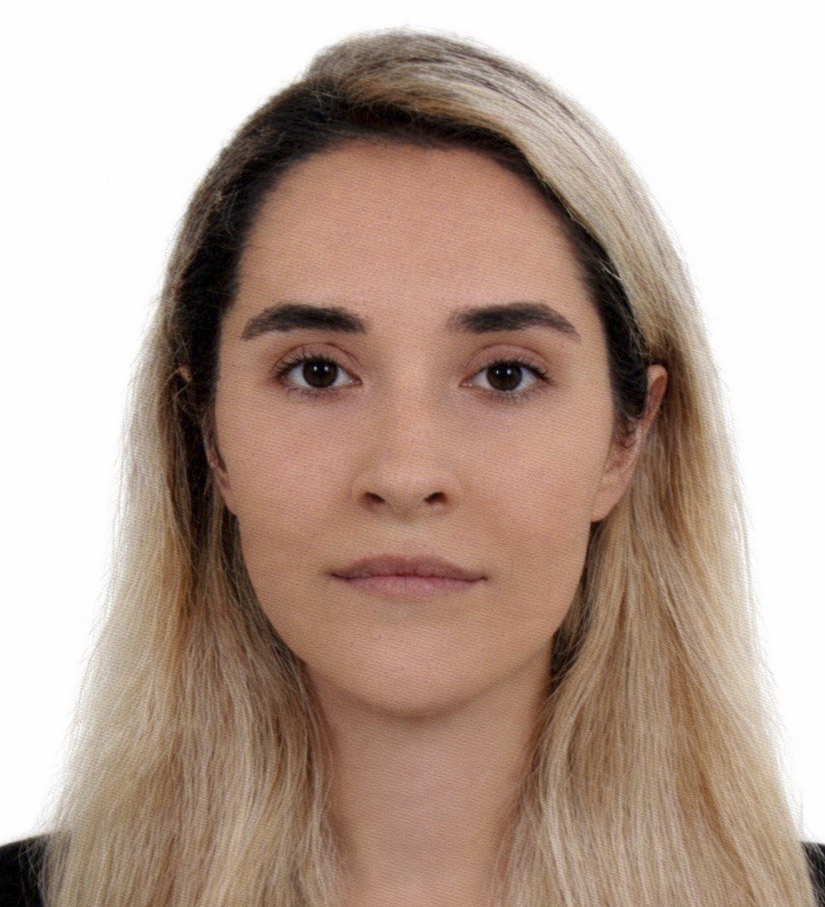}%
}%
}{\textbf{Parmida Geranmayeh} received the B.Sc. degree in Computer Engineering (Software Engineering) and the M.Sc. degree in Information Technology Engineering from the Iran University of Science and Technology (IUST), Tehran, Iran, in 2018 and 2020, respectively. She received the Dr.-Ing. degree in Computer Engineering from Humboldt University of Berlin, Germany, in 2025. She is currently a Postdoctoral Researcher with the Department of Communications Engineering (Nachrichtentechnik), Technical University of Dortmund, Germany. Her research interests include machine learning, reinforcement learning, optimization, beamforming, and 5G/6G wireless communication systems. She serves as a reviewer for several international journals, including \emph{Physical Communication} (Elsevier), \emph{Scientific Reports} (Springer Nature), and the \emph{International Journal of Communication Systems} (Wiley). She served as a Session Chair at the IEEE-APS Topical Conference on Antennas and Propagation in Wireless Communications in 2024 and received the Best Presenter Award at the IEEE International Conference on Artificial Intelligence for Sustainable Innovation (AI-SI 2025).}
\bio{%
\raisebox{-0.35cm}{%
\includegraphics[width=3.5cm,height=5.3cm,clip,keepaspectratio]{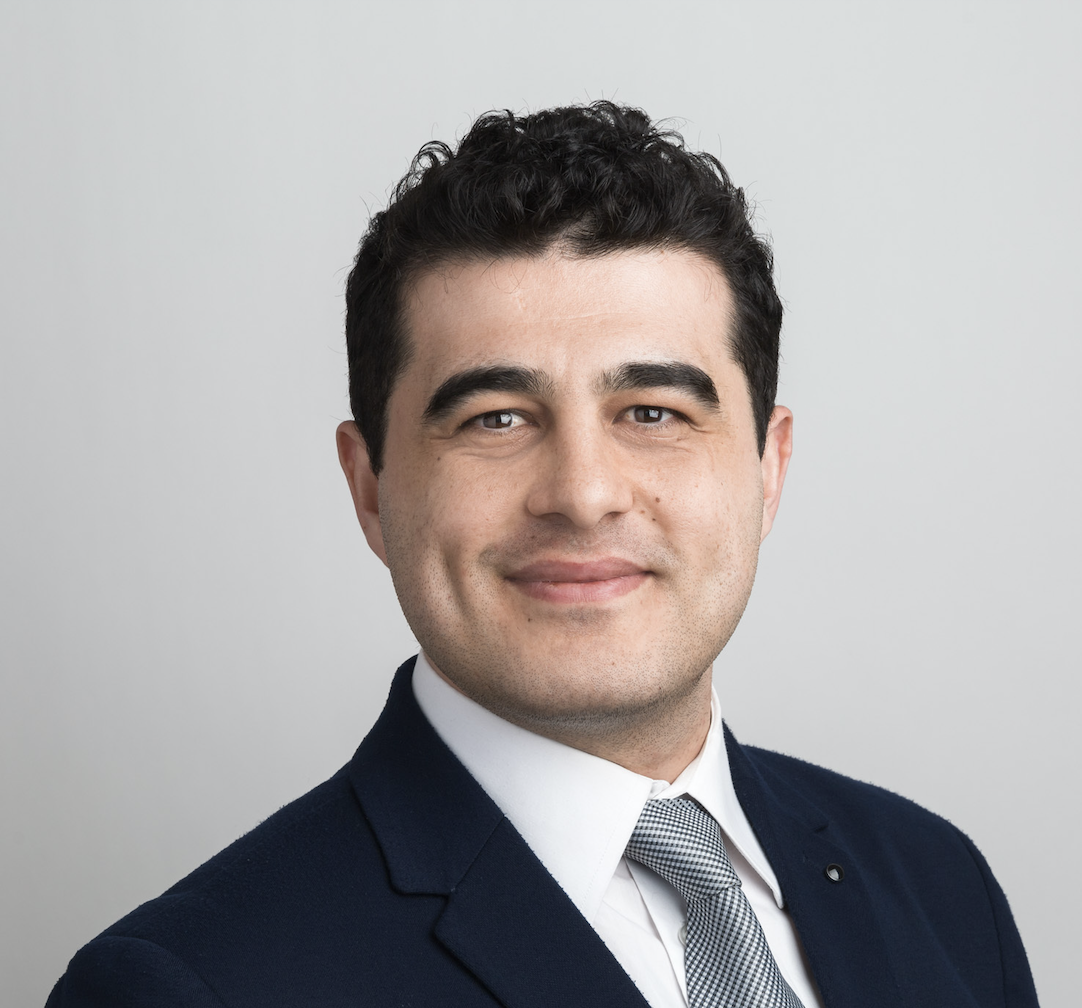}%
}%
}{\textbf{Onur Günlü} received the B.Sc. degree (Highest Distinction) in Electrical and Electronics Engineering from Bilkent University, Turkey in 2011; M.Sc. (Highest Distinction) and Dr.-Ing. (Ph.D. equivalent) degrees in Communications Engineering both from the Technical University of Munich (TUM), Germany in 2013 and 2018, respectively. He was a Working Student in the Communication Systems division of Intel Mobile Communications (IMC), now Apple Inc., in Munich, Germany during November 2012 - March 2013. Onur worked as a Research and Teaching Assistant at TUM Chair of Communications Engineering (LNT) between February 2014 - May 2019. As a Visiting Researcher, among more than twenty Research Stays at Top Universities and Companies, he was at TU Eindhoven, Netherlands during February 2018 - March 2018. Onur was a Visiting Research Group Leader at Georgia Institute of Technology, Atlanta, USA during February 2022 - March 2022. He was also a Visiting Professor at TU Dresden, Germany during February 2023 - March 2023. Following Research Associate and Group Leader positions at TUM, TU Berlin, and the University of Siegen, he joined Linköping University in October 2022 as an ELLIIT Assistant Professor and obtained tenure as an Associate Professor leading the Information Theory and Security Laboratory (ITSL) in August 2024. He became a Swedish Docent (Dr.-habil.) of Information Theory in December 2023 and an IEEE Senior Member in July 2024. Since September 2025, Onur has been a Tenured Full Professor leading the Lehrstuhl für Nachrichtentechnik (Institute of Communications Engineering ) at TU Dortmund (TUDO), Germany. He was also a Guest Professor at Linköping University, Sweden and since September 2026, he is a Guest Full Professor there. He has received the 2026 IEEE Communications Society \& Information Theory Society - Joint Best Paper Award, 2025 IEEE Information Theory Society – Joy Thomas Tutorial Paper Award, the 2023 ZENITH Research and Career Development Award, 2021 IEEE Transactions on Communications - Exemplary Reviewer Award, and the prestigious 2021 VDE Information Technology Society (ITG) Johann-Philipp-Reis Award. His research interests include distributed function computation, information-theoretic privacy and security, coding theory, 6G integrated sensing and communication (ISAC), and private learning. Among his publications is the book \emph{Key Agreement with Physical Unclonable Functions and Biometric Identifiers} (Dr. Hut Verlag, 2019). He serves as Associate Editor for {IEEE JOURNAL ON SELECTED AREAS IN COMMUNICATIONS}, {IEEE TRANSACTIONS ON COMMUNICATIONS}, and {ENTROPY} Journal, and was recently an Associate Editor of {EURASIP JOURNAL ON WIRELESS COMMUNICATIONS AND NETWORKING} and a Guest Editor of {IEEE JOURNAL ON SELECTED AREAS IN INFORMATION THEORY}. He also serves as a Working Group Leader for EU COST Action 6G-PHYSEC on Intelligent and Resilient Systems and as an Elected Member of the IEEE Signal Processing Society - Information Theory and Forensics (IFS) Technical Committee (TC).}

\end{adjustwidth} 

\PublishersNote{}

\end{document}